\documentclass[a4paper,11pt]{article}
\usepackage{jcappub} % for details on the use of the package, please
\usepackage{graphicx}
\usepackage{amsmath,amssymb}
\usepackage{epstopdf}

\usepackage{tabularx}
\usepackage{mathtools}
\usepackage{pifont}
\usepackage[colorinlistoftodos]{todonotes}
\usepackage[normalem]{ulem}
\usepackage[utf8]{inputenc}
\usepackage{bm}
\usepackage{subcaption}
\usepackage{booktabs}

\graphicspath{{Figures_gle/}}
\usepackage[utf8]{inputenc}

\begin{document}
	\title{Radial Oscillations and Stability of Neutron Stars with Antikaon Condensates}
%%%%%%%%%%%%%%%%%%%%%%%%%%%%%%%%%%%%%%%%%%%%%%%%%%%%%%%%%%%%%%%%%
\author[a]{Manisha Kumari,}
\affiliation[a]{Physics Group, Variable Energy Cyclotron Centre, Kolkata 700064, India}
\affiliation[b]{Homi Bhabha National Institute, Training School Complex, Anushakti Nagar, Mumbai-400085, India}
\emailAdd{maniyadav93@gmail.com}

\author[a,1]{Sujan Kumar Roy, \note{Corresponding author.}}
\emailAdd{sujan.kr@vecc.gov.in}

\author[a,b]{Soumen Podder,}
\emailAdd{s.podder@vecc.gov.in}

\author[a,b]{Suman Pal,}
\emailAdd{sumanvecc@gmail.com}

\author[a,b]{Gargi Chaudhuri}
\emailAdd{gargi@vecc.gov.in}

\def\be{\begin{equation}}
	\def\ee{\end{equation}}
\def\bearr{\begin{eqnarray}}
	\def\eearr{\end{eqnarray}}
\def\zbf#1{{\bf {#1}}}
\def\bfm#1{\mbox{\boldmath $#1$}}
\def\hf{\frac{1}{2}}
\def\kp{\zbf k+\frac{\zbf q}{2}}
\def\km{-\zbf k+\frac{\zbf q}{2}}
\def\hwo{\hat\omega_1}
\def\hwt{\hat\omega_2}

\abstract{Radial oscillations provide a direct probe of the stability and compressibility of neutron stars and are highly sensitive to the equation of state of dense matter. In this work, we investigate the impact of antikaon condensates on the radial oscillation properties of neutron stars. We model neutron star matter using equations of state with a wide range of stiffness. For this purpose, both non-linear and density-dependent relativistic mean-field frameworks are employed to develop equations of state that are consistent with current astrophysical constraints. We further consider the emergence of antikaon condensates ($K^-$ and $\bar{K}^0$) in the stellar core, which modifies the pressure--energy density relation of dense matter. We find that the nature of the transition from nuclear matter to the condensed phase is sensitive to the antikaon optical potential depth and underlying equation of state. We compute the fundamental and higher-order radial oscillation modes for neutron stars containing antikaon condensates over a range of antikaon optical potential depths. Our results demonstrate that the antikaon optical potential depth plays a decisive role in governing the systematic shifts observed in the radial oscillation frequencies, while also significantly reducing the stability limits and maximum masses of neutron stars. These imprints of antikaon condensation on radial oscillation spectra provide a promising avenue for future multi-messenger observations and high-frequency gravitational-wave searches to directly probe and constrain the internal composition and equation of state of neutron stars.}

\maketitle
\flushbottom
\section{Introduction}
\label{intro}

Neutron stars (NSs) are among the densest known objects in the universe, formed from the collapsed cores of massive stars following supernova explosions \citep{Baade:1934wuu}. Although their radii are only about 10--15 km, NSs can possess masses up to roughly twice that of the Sun, resulting in interior densities far exceeding nuclear saturation density. Since such extreme conditions cannot be reproduced in laboratory experiments, NSs provide a unique astrophysical setting for investigating the properties of cold, dense nuclear matter. The structure and observable characteristics of NSs are determined primarily by the equation of state (EoS) of dense matter. While the EoS is relatively well constrained near nuclear saturation density, its behavior at higher densities remains poorly understood. In this regime, new and exotic degrees of freedom--such as hyperons \citep{1996cost.book.....G, Bombaci:2016xzl}, meson condensates \citep{Mannarelli:2019hgn, ABMigdal_1972}, or deconfined quark matter \citep{Rajagopal:2000wf, Alford:2007xm}--may appear. The inclusion of these components typically softens the EoS, leading to notable changes in NS masses, radii, and stability. Consequently, identifying observational signatures of such exotic phases remains a key challenge in NS physics.

The observation of gravitational waves from the binary NS merger GW170817 marked a major advance in the study of matter under extreme conditions \citep{LIGOScientific:2017vwq, LIGOScientific:2018cki}. Such detections have established gravitational-wave astronomy as a powerful tool for probing the internal structure of NSs, providing strong constraints on the EoS and the composition of NSs \citep{LIGOScientific:2017vwq, LIGOScientific:2018cki, Most:2018hfd, Fattoyev:2017jql}. In addition to merger signals, NSs can reveal their internal properties through stellar oscillations. These oscillations are broadly classified into non-radial and radial modes \citep{Chandrasekhar:1964zz}. Non-radial modes involve angular distortions of the star and are efficient emitters of gravitational waves, making them particularly relevant for future detectors. On the other hand, radial oscillations are spherically symmetric and do not emit gravitational waves directly, which makes their detection inherently challenging. However, radial modes can couple to non-radial oscillations and thereby enhance gravitational-wave emission, particularly in highly dynamical environments such as NS-NS mergers. In the post-merger phase, a hypermassive NS may form, accompanied by a short gamma-ray burst and exhibiting high-frequency oscillations in the 1--4 kHz range, where the dynamics may be influenced by radial modes \citep{Chirenti:2019sxw}. Present gravitational-wave detectors, such as LIGO, are not sensitive enough in the kilohertz frequency range to detect the radial oscillations. Future third-generation observatories, including the Cosmic Explorer and the Einstein Telescope, are expected to improve sensitivity by roughly an order of magnitude, potentially enabling such observations \citep{Punturo:2010zz, Evans:2021gyd, Kalogera:2021bya}. Nevertheless, radial modes are of central importance because their frequencies are directly tied to stellar stability and the compressibility of dense matter, making them highly sensitive to the underlying EoS. In realistic astrophysical environments such as NS-NS mergers, core-collapse supernovae, and starquakes, radial oscillation modes can be excited and, through their coupling with non-radial modes, may produce observable indirect signatures in gravitational-wave emission.

The theoretical foundation of radial oscillations in relativistic stars was laid by Chandrasekhar \citep{Chandrasekhar:1964zz, Chandrasekhar:1964zza} and subsequently extended through detailed studies of radial pulsations in cold NSs \citep{Chanmugam_1977, Glass_1983, Vaeth_1992, Harrison_1965} and proto-NSs \citep{Gondek:1997fd}. Recent studies have shown that radial oscillation frequencies are highly sensitive to the microscopic composition of dense matter and can thus serve as effective probes of exotic phases in NS interiors. In particular, they have been used to investigate the presence of dark matter, deconfined quark matter, hadronic matter containing hyperons and $\Delta$ baryons, as well as phase transitions occurring at higher densities \citep{Kokkotas:2000up, Panotopoulos:2018ipq, Sagun:2020qvc, Sun:2021cez, Sen:2022kva, Li:2022qql, Routaray:2022acz, Routaray:2022utr, Rather:2023dom, Rather:2023tly, Rather:2024hmo, Jyothilakshmi:2025wru}. Motivated by these developments, radial oscillations provide a natural framework for examining the impact of phase transitions occurring in the dense cores of NSs.

Several types of phase transitions may arise in NS matter at high densities, including pion or kaon condensation and transitions to deconfined quark matter. In this work, we focus on the Bose--Einstein condensation of antikaons, which has been identified as a particularly favorable strange degree of freedom in dense matter \citep{Schaffner:1996th, Glendenning:1997ak, Schaffner-Bielich:1999fyk}. The possibility of $K^-$ condensation was first proposed by Kaplan and Nelson in the context of dense hadronic systems \citep{Kaplan:1986yq}. In NS matter, negatively charged antikaons may appear when the lepton chemical potential exceeds the in-medium energy of the $K^-$, allowing them to replace leptons as carriers of negative charge. This mechanism leads to a substantial softening of the EoS and can strongly influence both the global properties and the stability of NSs. The role of antikaon condensation in NSs and proto-NSs has been investigated in earlier studies exploring a limited zone in the pressure--energy density plane \citep{Lalazissis:1996rd, Pal:2000pb, Fattoyev:2010mx, Pons:2000iy, Pons:2000xf, Banik:2000dx, Banik:2001yw, Banik:2002qu, Gupta:2013rba, Char:2014cja, Batra:2017mfv, Banik:2014rga, Kundu:2022nva}. References~\citep{Thapa:2020usm, Thapa:2021kfo} examine the potential onset of antikaon condensation in $\beta$-equilibrated nuclear matter as well as in hyperonic matter containing $\Delta$ resonances.

%These studies are carried out within the framework of non-linear (NL) and density-dependent (DD) RMF models. In most investigations, antikaons are incorporated via a minimal coupling scheme \citep{Glendenning:1997ak}. Extended RMF models are typically implemented either by introducing non-linear (NL) self-interaction terms for the meson fields in the original Walecka model \citep{Serot:1984ey}, or by employing density-dependent (DD) meson--baryon couplings in a linear RMF formulation \citep{Typel:1999yq}.

While radial oscillations of NSs have been studied for purely nucleonic matter and for NS matter containing hyperons or deconfined quark matter, a systematic analysis of radial oscillations in NSs with antikaon condensates--covering a wide range of stiffness--has not yet been carried out. For this purpose, we consider both non-linear (NL) and density-dependent (DD) RMF approaches to develope EoSs with varying stifnesses. NL RMF models, such as NL3 \citep{Lalazissis:1996rd}, GMT \citep{Pal:2000pb}, and IUFSU \citep{Fattoyev:2010mx}, include meson self-interaction terms that control the stiffness of the EoSs \citep{Serot:1984ey}. It is important to note that these models have been widely used in the studies of nuclear matter and NSs. On the other hand, DD-RMF models, including DD2 \citep{Typel:2009sy}, PKDD \citep{Long:2003dn}, and DDME$\delta$ \citep{Roca-Maza:2011alv}, offer an alternative description in which meson--nucleon couplings vary explicitly with density, allowing for a more flexible and microscopically motivated treatment of nuclear interactions at high densities \citep{Typel:1999yq}. Consequently, we incorporate the antikaon condensates, study the radial oscillation properties of the resulting NSs and present a systematic comparison between different NS models.

% A systematic comparison between these two modeling approaches is therefore crucial for assessing uncertainties in the EoS, especially in the presence of exotic degrees of freedom like antikaon condensates, and for understanding their impact on the dynamical properties of NSs.

% The threshold density for antikaon condensation is known to be highly sensitive to the depth of the antikaon optical potential, with deeper potentials leading to an earlier onset of $K^-$ condensation and enhanced softening of the EoS \citep{Banik:2000dx, Banik:2002qu}. One of the primary objectives of this study is therefore to determine the range of antikaon optical potentials that remains compatible with current observational constraints on the maximum NS mass.

The paper is organized as follows. In sections~\ref{sec:2_A} and \ref{sec:2_B}, we describe the theoretical framework used to model NSs with antikaon condensation, including details of the NL and DD-RMF parameterizations. Section~\ref{sec:2_C} introduces the equations governing NS structure and radial oscillations. In section~\ref{sec:3}, we present and discuss the resulting EoSs, mass--radius relations, radial oscillation frequencies, and corresponding eigenfunctions for different values of the antikaon optical potential. Finally, section~\ref{sec:4} summarizes our main results and outlines their implications for NS stability and for future gravitational-wave detections and multi-messenger studies of dense matter.

\section{Methodology}
\label{sec:2}                                                                   

\subsection{Relativistic Mean-Field Model} \label{sec:2_A}

To investigate radial oscillations in NS matter containing antikaon condensates, we employ both NL and DD-RMF frameworks. The matter composition includes nucleons ($N \equiv  p, n$), antikaons ($\bar{K}\equiv K^-, \bar{K}^0$), and leptons ($l \equiv e^-, \mu^-$).
 The total Lagrangian density describing the interacting system is written as \citep{Glendenning:1997ak, Pal:2000pb, Banik:2001yw, Banik:2000dx}
%---------------------------------------------------------------------------
\begin{eqnarray}\label{Eqs.1}
	\begin{aligned}
		\mathcal{L} & = \sum_{N} \bar{\psi}_N(i\gamma_{\mu} D^{\mu} - m^{*}_N) \psi_N   + D^*_\mu \bar{K} D^\mu K - m^{*^2}_{\bar{K}} \bar{K} K + \frac{1}{2}(\partial_{\mu}\sigma\partial^{\mu}\sigma  - m_{\sigma}^2 \sigma^2) - \frac{\kappa}{3!} (g_{\sigma N} \sigma)^3 \\
		&   - \frac{\lambda}{4!} (g_{\sigma N} \sigma)^4 + \frac{1}{2}(\partial_{\mu}\delta\partial^{\mu}\delta - m_{\delta}^2 \delta^2) -  \frac{1}{4}\omega_{\mu\nu}\omega^{\mu\nu} + \frac{1}{2}m_{\omega}^2\omega_{\mu}\omega^{\mu} + \frac{\zeta}{4!} (g^2_{\omega N} \omega_\mu \omega^\mu)^2 \\ 
		&  - \frac{1}{4}\boldsymbol{G}_{\mu\nu} \cdot \boldsymbol{G}^{\mu\nu} + \frac{1}{2}m_{\rho}^2\boldsymbol{\rho}_{\mu} \cdot \boldsymbol{\rho}^{\mu}  + \Lambda_{\rm v} (g^2_{\rho N} \boldsymbol{\rho}_\mu \boldsymbol{\rho}^\mu) (g^2_{\omega N} \omega_\mu \omega^\mu) + \sum_{l} \bar{\psi}_l (i\gamma_{\mu} \partial^{\mu} - m_l)\psi_l.
	\end{aligned}
\end{eqnarray}
%---------------------------------------------------------------------------
The corresponding field strength tensors for the vector mesons are defined as
%-----------------------------------------------------
\begin{equation} \label{Eqs.2}
	%-----------------------------------------------------
	\begin{aligned}
		\omega_{\mu \nu} & = \partial_{\mu}\omega_{\nu} - \partial_{\mu}\omega_{\nu} ,\\
		\boldsymbol{G}_{\mu \nu} & = \partial_{\nu}
		\boldsymbol{\rho}_{\mu} - \partial_{\mu}\boldsymbol{\rho}_{\nu}.
	\end{aligned}    
\end{equation}
%-----------------------------------------------------
The covariant derivative appearing in Eq.\eqref{Eqs.1} is given by
%-----------------------------------------------------
\begin{equation} \label{Eqs.3}
	D_\mu = \partial_\mu + ig_{\omega j} \omega_\mu + ig_{\rho j} \boldsymbol{\tau}_j \cdot \boldsymbol{\rho}_{\mu}.   
\end{equation}
%-----------------------------------------------------

The strong interactions between nucleons and kaons are mediated by the isoscalar-scalar meson $\sigma$, isovector scalar meson $\delta$, isoscalar vector meson $\omega$ and isovector vector meson $\rho$, each with thier respective masses and coupling constants. The  $\boldsymbol{\tau}$ represents the isospin operator for the isovector ($\delta$ and $\rho$) fields. The NL self-interaction terms of the $\sigma$ and $\omega$ mesons, governed by the parameters $\kappa$, $\lambda$, and $\zeta$, are essential for reproducing realistic properties of symmetric nuclear matter. The mixed $\omega$--$\rho$ coupling, controlled by $\Lambda_{\rm v}$, regulates the density dependence of the symmetry energy. The effects associated with the $\delta$ meson field are taken into account exclusively within the DDME$\delta$ parametrization.
Within the mean-field approximation, the effective masses of nucleons and antikaons are modified due to their interactions with the scalar mesons and are given by
%----------------------------------------------------------------------
\begin{equation} \label{Eqs.4}
	\begin{aligned}
		m_{N}^* = m_N - g_{\sigma N}\sigma - g_{\delta N}\tau_{3N}\delta,  \\ m_{\bar{K}}^* = m_{\bar{K}} - g_{\sigma K}\sigma  - g_{\delta K}\tau_{3\bar{K}}\delta.
	\end{aligned}
\end{equation}
%----------------------------------------------------------------------
Here, $m_N$ and $m_{\bar{K}}$ denote the bare masses of nucleons and antikaons, respectively. The antikaon isospin doublet is $\bar{K} = (K^-, \bar{K}^0)$ with isospin projections $(-1/2, +1/2)$.

The mean-field equations of motion for the meson fields are obtained by minimizing the Lagrangian density and can be expressed as
%----------------------------------------------------------------------
\begin{equation} \label{Eqs.5}
	\begin{aligned}
		\sigma & = -\frac{\kappa}{2m_{\sigma}^2} g^3_{\sigma N} \sigma^2 -\frac{\lambda}{6m_{\sigma}^2} g^4_{\sigma N} \sigma^3 + \sum_{N} \frac{1}{m_{\sigma}^2} g_{\sigma N}\rho_{N}^s + \sum_{\bar{K}} \frac{1}{m_{\sigma}^2} g_{\sigma K}\rho_{\bar{K}},\\
		\omega & = -\frac{\zeta}{2m_{\omega}^2} g^4_{\omega N} \omega^3 -\frac{2\Lambda_{\rm v}}{m_{\omega}^2} g^2_{\rho N} g^2_{\omega N} \rho^2 \omega + \sum_{N} \frac{1}{m_{\omega}^2} g_{\omega N}\rho_{N} - \sum_{\bar{K}} \frac{1}{m_{\omega}^2} g_{\omega K}\rho_{\bar{K}}, \\
		\rho & = -\frac{2\Lambda_{\rm v}}{m_{\rho}^2} g^2_{\rho N} g^2_{\omega N} \omega^2 \rho  + \sum_{N} \frac{1}{m_{\rho}^2} g_{\rho N}
		\boldsymbol{\tau}_{3N}\rho_{N} + \sum_{\bar{K}} \frac{1}{m_{\rho}^2} g_{\rho K}
		\tau_{3\bar{K}}\rho_{\bar{K}}, \\
		\delta & =  \frac{1}{m_{\delta}^2} g_{\delta N}\tau_{3N}\rho_{N}^s + \sum_{\bar{K}} \frac{1}{m_{\delta}^2} g_{\delta K}\tau_{3\bar{K}} \rho_{\bar{K}}.
	\end{aligned}    
\end{equation}
%----------------------------------------------------------------------

In the IUFSU parametrization, the NL coefficients $\kappa$, $\lambda$, $\zeta$, and the mixed coupling $\Lambda_{\rm v}$ take nonzero values \citep{Fattoyev:2010mx}. In contrast, the NL3 \citep{Lalazissis:1996rd} and GMT \citep{Pal:2000pb} models exclude the $\zeta$ self-interaction of the $\omega$ meson as well as the $\omega$--$\rho$ mixing term governed by $\Lambda_{\rm v}$.
For DD-RMF models, neither the NL isoscalar meson self-interaction terms nor the $\omega$--$\rho$ cross-coupling contribution is included \citep{Hofmann:2000vz,Hofmann:2000mc}. The scalar and vector (number) densities of nucleon at zero temperature are defined as 
%----------------------------------------------------------------------
\begin{equation} \label{Eqs.6}
	\begin{aligned}
		\rho_{N}^s&= \langle\bar{\psi}_N \psi_N \rangle  = \frac{m^{*}_{N}}{2 \pi^2} \left[ p_{{F}_{N}} E_{F_N}  - m_{N}^{*^2} \ln \left( \frac{p_{{F}_N} + E_{F_N}}{m_{N}^{*}} \right) \right], \\
		\rho_{N}&=\langle\bar{\psi}_N \gamma^0 \psi_N\rangle  = \frac{p_{{F}_{N}}^{3}}{3 \pi^2},
	\end{aligned}    
\end{equation}
%----------------------------------------------------------------------
where $p_{F_N}$ and $E_{F_N}$ denote the Fermi momentum and Fermi energy of the nucleon species $N$.
For s-wave antikaon condensation, the number density of antikaons is expressed as \citep{Glendenning:1997ak}
%----------------------------------------------------------------------
\begin{equation} \label{Eqs.7}
	\begin{aligned}
		\rho_{\bar{K}} & = 2 \left( \omega_{\bar{K}} + g_{\omega K} \omega - g_{\rho K}\tau_{3\bar{K}} \rho \right)  = 2 m^*_K \bar{K} K
	\end{aligned}
\end{equation}
%----------------------------------------------------------------------
The in-medium energies of antikaons are given by
%----------------------------------------------------------------------
\begin{equation} \label{Eqs.8}
	\omega_{\bar{K}} = m^*_{\bar{K}} - g_{\omega K} \omega + g_{\rho K} \tau_{3\bar{K}} \rho.
\end{equation}
%----------------------------------------------------------------------
The chemical potential of nucleons takes the form
%----------------------------------------------------------------------
\begin{equation} \label{Eqs.9}
	\begin{aligned}
		& \mu_{N} = \sqrt{p_{F_N}^2 + m_{N}^{*2}} + g_{\omega N}\omega + g_{\rho N} \boldsymbol{\tau}_{3N} \rho + \Sigma^{r},
	\end{aligned}
\end{equation}
%----------------------------------------------------------------------
where the rearrangement term $\Sigma^r$ arises only in  RMF models and guarantees thermodynamic consistency.
%----------------------------------------------------------------------
\begin{equation}\label{Eqs.10}
	\begin{aligned}
		\Sigma^{r} & = \sum_{N} \left[ \frac{\partial g_{\omega N}}{\partial \rho_B}\omega \rho_{N} - \frac{\partial g_{\sigma N}}{\partial \rho_B} \sigma \rho_{N}^s + \frac{\partial g_{\rho N}}{\partial \rho_B} \rho  \boldsymbol{\tau}_{3N} \rho_{N} - \frac{\partial g_{\delta N}}{\partial \rho_B} \delta  \boldsymbol{\tau}_{3N} \rho_{N}^s \right],
	\end{aligned}
\end{equation}
%----------------------------------------------------------------------
where $\rho_B= \sum_{N} \rho_N$ is the total baryon number density. This rearrangement contribution affects the thermodynamic description solely through its role in the pressure. Such a term does not arise in NL-RMF models and is therefore omitted in that framework.
The total energy density from the NS matter is written by
%----------------------------------------------------------------------
%\begin{widetext}
	\begin{eqnarray} \label{Eqs.11}
		\begin{aligned}
			\varepsilon_f & = \frac{1}{2}m_{\sigma}^2 \sigma^{2} + \frac{1}{2}m_{\delta}^2 \delta^{2} + \frac{1}{2} m_{\omega}^2 \omega^2 + \frac{1}{2}m_{\rho}^2 \rho^2 \\
			& + \sum_N \frac{1}{\pi^2} \left[ p_{{F}_N} E^3_{F_N} - \frac{m_{N}^{*2}}{8} \left( p_{{F}_N} E_{F_N} + m_{N}^{*2} \ln \left( \frac{p_{{F}_N} + E_{F_N}}{m_{N}^{*}} \right) \right) \right]  \\ & + \frac{1}{\pi^2}\sum_l \left[ p_{{F}_l} E^3_{F_l} - \frac{m_{l}^{2}}{8} \left( p_{{F}_l} E_{F_l} + m_{l}^{2} \ln \left( \frac{p_{{F}_l} + E_{F_l}}{m_{l}} \right) \right) \right] .			
		\end{aligned}
	\end{eqnarray}
%\end{widetext}
%----------------------------------------------------------------------
The presence of antikaon condensates introduces an additional contribution to the energy density of the system, which is expressed as
%----------------------------------------------------------------------
\begin{equation} \label{Eqs.12}
	\varepsilon_{\bar{K}} = m^*_{\bar{K}} \rho_{\bar{K}}
\end{equation}
%----------------------------------------------------------------------
 Accordingly, the total energy density of NS matter is obtained by adding the fermionic and antikaonic components, $\varepsilon = \varepsilon_f + \varepsilon_{\bar{K}}$. Since antikaons form a Bose--Einstein condensate, they do not contribute directly to the pressure. The pressure is therefore determined entirely by the fermionic sector and is calculated using the thermodynamic relation
 %----------------------------------------------------------------------
\begin{equation} \label{Eqs.13}
	p = \sum_{N} \mu_N \rho_N + \sum_{l} \mu_l \rho_l - \varepsilon_f.
\end{equation}
%----------------------------------------------------------------------

At low baryon densities, the matter composition is restricted to neutrons, protons, and electrons, which satisfy both charge neutrality and $\beta$-equilibrium. Under these conditions, chemical equilibrium is ensured through the relation
%----------------------------------------------------------------------
\begin{equation} \label{Eqs.14}
	\mu_n = \mu_p + \mu_e
\end{equation}
%----------------------------------------------------------------------
As the density increases, the electron chemical potential grows and eventually reaches the muon rest mass, at which point muons begin to populate the system. The onset of muons is therefore determined by the condition $\mu_e = m_\mu$.

At still higher densities, weak interaction processes involving strangeness become energetically allowed within the NS core \citep{Prakash:1996xs,1996cost.book.....G}. These processes, such as $N \leftrightarrow N + \bar{K}$ and $e^- \leftrightarrow K^-$, impose additional chemical equilibrium constraints. As a consequence, antikaon condensation sets in when the following threshold conditions are satisfied:
%----------------------------------------------------------------------
\begin{equation} \label{Eqs.15}
	\begin{aligned}
		\mu_n - \mu_p = \omega_{K^-} = \mu_e, \quad \omega_{\bar{K}^0} = 0
	\end{aligned}
\end{equation}
%----------------------------------------------------------------------
The charge neutrality conditions in the nucleonic and kaon condensed phases are given by
%----------------------------------------------------------------------
\begin{equation} \label{Eqs.16}
	\begin{aligned}
		\rho_p - \rho_e - \rho_\mu = 0, \\
		\rho_p - \rho_e - \rho_\mu - \rho_{K^-}= 0. 
        \end{aligned}
\end{equation}
%----------------------------------------------------------------------

\subsection{Coupling parameters} \label{sec:2_B}

To specify the meson--nucleon interaction strengths, several established RMF parametrizations are employed. Within the NL-RMF framework, the NL3 \citep{Lalazissis:1996rd}, GMT \citep{Pal:2000pb}, and IUFSU \citep{Fattoyev:2010mx} parameter sets are adopted for the meson--nucleon couplings. For the  RMF approach, the DD2 \citep{Typel:2009sy}, PKDD \citep{Long:2003dn} , and DDME$\delta$ \citep{Roca-Maza:2011alv} parametrizations are used. The corresponding meson masses and coupling constants for both NL and DD-RMF models are summarized in table~\ref{tab:1}.

In all NL-RMF calculations, the bare nucleon mass is fixed at $m_N = 939$ MeV. For the DD-RMF models, slightly different bare nucleon masses are employed, namely $m_N = 939$ MeV for DD2, $938.565$ MeV for PKDD, and $939.573$ MeV for DDME$\delta$. The vacuum masses of the antikaons are taken as $493.69$ MeV for $K^-$ and $497.611$ MeV for $\bar{K}^0$ throughout this work.
%----------------------------------------------------------------------
\begin{table} [ht!]
	\centering
	\caption{Meson masses (in MeV) and corresponding meson--nucleon coupling constants for the NL and DD-RMF parametrizations employed in this work, evaluated at nuclear saturation density $\rho_0$.}
	\begin{tabular}{lccccc}
        \hline \hline
        Parameter & NL3 & GMT & IUFSU & DD2 & PKDD \\
        \hline
        $m_\sigma$ (MeV) & 508.194 & 511.198 & 491.500 & 550.124 & 555.5112 \\
        $m_\omega$ (MeV) & 782.501 & 783.000 & 782.500 & 783.000 & 783.000 \\
        $m_\rho$ (MeV) & 763.00 & 770.00 & 763.00 & 763.00 & 763.00 \\
        $g^2_{\sigma N}$ & 104.3871 & 98.8036 & 99.4266 & 114.2052 & 115.3159 \\
        $g^2_{\omega N}$ & 80.0667 & 86.0367 & 184.6877 & 178.0186 & 172.8594 \\
        $g^2_{\rho N}$ & 165.5854 & 151.2433 & 169.8349 & 52.6188 & 73.9531 \\
        $\kappa$ (MeV) & 3.85929 & 4.24804 & 3.3808 & -- & -- \\
        $\lambda$ & $-0.015906$ & $-0.014868$ & 0.000296 & -- & -- \\
        $\zeta$ & -- & -- & 0.03 & -- & -- \\
        $\Lambda_{\rm v}$ & -- & -- & 0.046 & -- & -- \\
        \hline
	\end{tabular}
	\label{tab:1}
\end{table}
%----------------------------------------------------------------------
Within the DDME$\delta$ model, the meson--nucleon coupling constants explicitly depend on the baryon density and are parameterized as
%----------------------------------------------------------------------
\begin{equation} \label{Eqs.17}
	g_{i N}(\rho)=g_{i N}(\rho_{0})f_{i}(x)\quad\mathcal{for}\quad
	i=\sigma,\omega,\delta,\rho
\end{equation}
%----------------------------------------------------------------------
where $x = \rho/\rho_0$ and $\rho_0$ denotes the saturation density of symmetric nuclear matter. The density dependence is encoded in the function
%----------------------------------------------------------------------
\begin{equation} \label{Eqs.18}
	f_{i}(x)=a_{i}\frac{1+b_{i}(x+d_{i})^{2}}{1+c_{i}(x+e_{i})^{2}} .
\end{equation}
%----------------------------------------------------------------------
For the DD2 and PKDD parametrizations, the density dependence of the $\sigma$- and $\omega$-meson couplings follows the same functional form as in the DDME$\delta$ model. However, the $\rho$-meson coupling is described by an exponential dependence given by \citep{Typel:2009sy}
%----------------------------------------------------------------------
\begin{equation} \label{Eqs.19}
	g_{\rho N}(n)= g_{\rho N}(n_{0}) e^{-a_{\rho}(x-1)}.
\end{equation}
%----------------------------------------------------------------------
The coefficients governing the density dependence in equations~(\ref{Eqs.18})--(\ref{Eqs.19}) are determined using fitting procedures described in Refs.~\citep{Typel:2009sy,PhysRevC.71.024312,Long:2003dn,Roca-Maza:2011alv}. The numerical values of these parameters for the DD2, PKDD, and DDME$\delta$ models are listed in tables~\ref{tab:2} and \ref{tab:3}.
%----------------------------------------------------------------------
\begin{table*}[ht!]
	\centering
	\caption{Values of the DD coupling coefficients for the DD2 and PKDD parametrizations at saturation density $\rho_0$.}
	\begin{tabular}{ccccccc}
		\hline \hline
		Model & $i$ & $a_i$ & $b_i$ & $c_i$ & $d_i$ & $e_i$ \\
		\hline
		& $\sigma$ & 1.3576  & 0.63444 & 1.00536 & 0.5758  & 0.5758 \\
		DD2
		& $\omega$ & 1.36972 & 0.49647 & 0.81775 & 0.63845 & 0.63845 \\
		& $\rho$   & 0.51890 & --      & --      & --      & --      \\
		\hline
		& $\sigma$ & 1.32742 & 0.43513 & 0.69167 & 0.69421 & 0.69421 \\
		PKDD
		& $\omega$ & 1.34217 & 0.37117 & 0.61140 & 0.73838 & 0.73838 \\
		& $\rho$   & 0.18330 & --      & --      & --      & --      \\
		\hline
	\end{tabular}
	\label{tab:2}
\end{table*}
%----------------------------------------------------------------------
%----------------------------------------------------------------------
\begin{table*} [ht!]
	\centering
	\caption{Meson masses, meson--nucleon coupling strengths at saturation density, and parameters governing the density dependence of the coupling functions for the DDME$\delta$ model.}
	\begin{tabular}{cccccccc}
		\hline \hline
		$i$ & $m_{i}$ (MeV) & $g_{i}$ ($\rho_0$) & $a_{i}$ & $b_{i}$ & $c_{i}$ & $d_{i}$ & $e_{i}$ \\
		\hline
	$\sigma$ & 566.1577 & 10.3325 & 1.3927 & 0.1901 & 0.3679 & 0.9519 & 0.9519 \\
	$\omega$ & 783.0000 & 12.2904 & 1.4089 & 0.1698 & 0.3429 & 0.9860 & 0.9860 \\
	$\delta$ & 983.0000 & 14.3040 & 1.5178 & 0.3262 & 0.6041 & 0.4257 & 0.5885  \\
	$\rho$ & 763.0000 & 12.6256 & 1.8877 & 0.0651 & 0.3469 & 0.9417 & 0.9737  \\
	\hline
	\end{tabular}
	\label{tab:3}
\end{table*} 
%----------------------------------------------------------------------

All parametrizations considered in this work reproduce standard nuclear matter properties at saturation density $\rho_0$, as summarized in table~\ref{tab:4}. The quantities $E/A$, $K_0$, $a_{\rm sym}$, and $m_N^*/m_N$ denote the binding energy per nucleon, incompressibility modulus, symmetry energy coefficient, and the ratio of effective to bare nucleon mass, respectively.
%----------------------------------------------------------------------
\begin{table} [ht!]
	\centering
	\caption{The nuclear properties of the NL and  RMF models at respective nuclear saturation densities.}
	\begin{tabular}{cccccc}
		\hline \hline
		Model & $\rho_0$ (fm$^{-3}$) & $E/A$ (MeV) & $K_0$ (MeV) & $a_{sym}$ (MeV) & $m^*_N/m_N$ \\
		\hline
		NL3 & 0.148 & $-16.299$ & 271.760 & 37.40 & 0.6000 \\
		GMT & 0.145 & $-16.300$ & 281.000 & 36.90  & 0.6340 \\
		IUFSU & 0.155 & $-16.400$ & 231.200 & 31.30  & 0.6000 \\
		DD2 & 0.149065 & $-16.020$ & 242.700 & 32.73  & 0.5625 \\
		PKDD & 0.149552 & $-16.267$ & 262.181 &  36.79  &  0.5712 \\
		DDME$\delta$ & 0.152 & $-16.120$ & 219.100 & 32.35  & 0.6090 \\
		\hline
	\end{tabular}
	\label{tab:4}
\end{table}
%----------------------------------------------------------------------

In the  DD-RMF framework, the antikaon--meson couplings are treated as density independent. The vector coupling constants are determined using quark model arguments and isospin counting rules \citep{Banik:2001yw,Schaffner:1996th,Glendenning:1997ak}, leading to
%----------------------------------------------------------------------
\begin{equation}\label{Eqs.20}
	g_{\omega K} = \frac{1}{3} g_{\omega N}, \quad  g_{\rho K} = g_{\rho N},
\end{equation}
%----------------------------------------------------------------------
The sigma meson-antikaon coupling is constrained by the real part of the $K^-$ optical potential in symmetric nuclear matter at saturation density, which is defined as \citep{Schaffner:1996th,Banik:2001yw,Char:2014cja}
%----------------------------------------------------------------------
\begin{equation} \label{Eqs.21}
	U_{\bar{K}} (\rho_0) = - g_{\sigma K} \sigma (\rho_0) - g_{\omega K} \omega (\rho_0) + \Sigma^r_{N} (\rho_0)
\end{equation}
%----------------------------------------------------------------------
Here, $\Sigma_N^r(\rho_0)$ represents the rearrangement contribution arising solely from nucleons and is absent in NL-RMF models. In this study, the antikaon optical potential is varied within the range of $-160 \leq U_{\bar{K}} \leq -120$ MeV and the corresponding $\sigma$ meson--antikaon coupling constants are presented in table~\ref{tab:5}.
%----------------------------------------------------------------------
\begin{table} [ht!]
	\centering
	\caption{calar $\sigma$-meson--antikaon coupling constants $g_{\sigma K}$ corresponding to $U_{\bar{K}}$ for NL3, GMT, IUFSU, DD2, PKDD and DDME$\delta$ parametrizations at saturation density, $\rho_0$.}
	\begin{tabular}{cccccc}
		\hline \hline
		Model & \multicolumn{5}{c}{$U_{\bar{K}}$ (MeV)} \\
		& $-120$ & $-130$ & $-140$ & $-150$ & $-160$ \\
		\hline
		NL3 & 0.4705 & 0.7395 & 1.0086 & 1.2776 & 1.5467 \\
		GMT & 0.8215 & 1.1109 & 1.4004 & 1.6898 & 1.9793 \\
		IUFSU & 0.5894 & 0.8606 & 1.13618 & 1.4031 & 1.6743 \\
		DD2 & 0.3152 & 0.5754 & 0.8356 & 1.0957 & 1.3559 \\
		PKDD & 0.4311 & 0.6975  & 0.9640 & 1.2305 & 1.4969 \\
        DDME$\delta$ & 0.4309 & 0.6930 & 0.9552  & 1.2173 & 1.7415 \\
		\hline
	\end{tabular}
	\label{tab:5}
\end{table}
%----------------------------------------------------------------------
Following the approach of Ref.~\citep{PhysRevC.82.055801}, the $\delta$ meson--antikaon coupling, $g_{\delta K}$ is taken to be 0.95$g_{\delta N}$. This choice ensures consistency with the corresponding nucleon--meson interaction strengths and allows for a realistic treatment of isovector-scalar effects in dense matter.

\subsection{Radial Oscillation} \label{sec:2_C}
\subsubsection{Static Background} \label{static}

In the context of general relativity, we consider spherically symmetric stellar configurations and treat radial motions as perturbations about a static equilibrium. The unperturbed NS background is described by the static, spherically symmetric spacetime metric in $(t,r,\theta,\phi)$ coordinates,
%----------------------------------------------------------------------
\begin{align}
    ds^2_{0} &= - e^{\nu_{0}(r)} dt^2 + e^{\lambda_{0}(r)} dr^2 + r^2 d\Omega^2 ,
\label{eq_radial_1}
\end{align}
%----------------------------------------------------------------------
where $d\Omega^2$ denotes the metric on a unit two-sphere. The subscript $(0)$ labels quantities associated with the unperturbed configuration. The metric potentials $\nu_{0}(r)$ and $\lambda_{0}(r)$ are determined from the equations of hydrostatic equilibrium.

Setting $G=c=1$, the Einstein field equations for the unperturbed star,
%----------------------------------------------------------------------
\begin{align}
G_\zeta^{~\eta} = 8\pi T_\zeta^{~\eta},
\label{eq_radial_1_1}
\end{align}
%----------------------------------------------------------------------
yield the equilibrium equations for a static stellar configuration,
%----------------------------------------------------------------------
\begin{align}
 \frac{dM}{dr} &= 4 \pi r^2 \varepsilon_{0}, \nonumber\\
 \frac{dP_{0}}{dr} &= - \frac{\left( \varepsilon_{0} + P_{0} \right) \left( M + 4\pi r^3 P_{0} \right)}{r\left(r - 2M\right)}, \nonumber\\
 \frac{d\nu_{0}}{dr} &= -\frac{2}{\left( \varepsilon_{0} + P_{0} \right)} \frac{dP_{0}}{dr}, \nonumber\\
 e^{\lambda_{0}} &= \frac{1}{1-\frac{2M}{r}},
\label{eq_radial_2}
\end{align}
%----------------------------------------------------------------------
where $M(r)$ denotes the gravitational mass enclosed within radius $r$. The stellar surface is defined at $r=R$, where the pressure vanishes, $P_0(R)=0$. The first two equations constitute the Tolman--Oppenheimer--Volkoff (TOV) equations, which determine the mass--radius relation of the star \citep{Tolman_1939_wfrdC, Oppenheimer:1939ne}. The remaining equations determine the metric potentials for the unperturbed configuration. Here, $\varepsilon_0$ and $P_0$ denote the energy density and pressure of the static background.

\subsubsection{Radially Perturbed Stars}

We now introduce radial perturbations about the equilibrium configuration while preserving spherical symmetry. The spacetime metric of the perturbed star can be written as \citep{Chandrasekhar:1964zz}
%----------------------------------------------------------------------
\begin{align}
    ds^2 &= - e^{\nu(t,r)} dt^2 + e^{\lambda(t,r)} dr^2 + r^2 d\Omega^2 ,
\label{eq_radial_3}
\end{align}
%----------------------------------------------------------------------
where $\nu(t,r)$ and $\lambda(t,r)$ represent the perturbed metric functions. We decompose the physical variables as
%----------------------------------------------------------------------
\begin{align}
    \nu &= \nu_0 + \delta\nu, \quad & \lambda &= \lambda_0 + \delta\lambda, \nonumber\\
    \varepsilon &= \varepsilon_0 + \delta\varepsilon, \quad & P &= P_0 + \delta P,
\label{eq_radial_5}
\end{align}
%----------------------------------------------------------------------
with $\delta f$ denoting Eulerian perturbations in function $f$.

The radial motion of fluid elements induced by the perturbation is described by the four-velocity
%----------------------------------------------------------------------
\begin{align} 
    u^{\mu} &= \left[ e^{-\nu_{0}/2},\, e^{-\nu_{0}/2} \mathcal{V},\, 0,\, 0 \right], \label{eq_radial_4_1} \\ 
    \mathcal{V} &= \frac{d \Delta r}{dt},
    \label{eq_radial_4_2}
\end{align}
%----------------------------------------------------------------------
where $\Delta r(t,r)$ denotes the Lagrangian radial displacement.

Linearizing Eq.~\eqref{eq_radial_1_1} to first order in the perturbations yields the equations governing radial oscillations. Substituting Eq.~\eqref{eq_radial_4_1} into the energy--momentum tensor,
%----------------------------------------------------------------------
\begin{align}
T_\zeta^{~\eta} = \left( \varepsilon + P \right) u_\zeta u^{\eta} + P \delta_\zeta^{~\eta},
\label{eq_radial_8}
\end{align}
%----------------------------------------------------------------------
and linearizing the $G_0^{~0}$ component gives
%----------------------------------------------------------------------
\begin{align}
\frac{\partial}{\partial r}(r e^{-\lambda_0} \delta\lambda) = 8\pi r^2 \delta\varepsilon.
\label{eq_radial_6_1}
\end{align}
%----------------------------------------------------------------------

Similarly, linearizing the $G_1^{~1}$ component yields
%----------------------------------------------------------------------
\begin{align}
\frac{e^{-\lambda_0}}{r}\left( \frac{\partial}{\partial r}\delta\nu - \frac{d\nu_0}{dr}\delta\lambda \right) = \frac{e^{-\lambda_0}}{r^2}\delta\lambda + 8\pi \delta P.
\label{eq_radial_6_2}
\end{align}
%----------------------------------------------------------------------

The $G_0^{~1}$ component leads to
%----------------------------------------------------------------------
\begin{align}
\frac{e^{-\lambda_0}}{r} \delta \lambda &= -8\pi (\varepsilon_0 + P_0)\Delta r, \nonumber\\
\Rightarrow \delta \lambda &= -\Delta r \frac{d}{dr} \left( \nu_0 + \lambda_0 \right),
\label{eq_radial_6_3}
\end{align}
%----------------------------------------------------------------------
where we have used Eq.~\eqref{eq_radial_2} to express $\delta\lambda$ in terms of $\Delta r$ and the background metric functions.

Substituting Eq.~\eqref{eq_radial_6_3} into Eq.~\eqref{eq_radial_6_1} yields
%----------------------------------------------------------------------
\begin{align}
\delta \varepsilon = -\Delta r \frac{d\varepsilon_0}{dr} - \frac{e^{\nu_0/2}}{r^2}(\varepsilon_0 + P_0) \frac{\partial}{\partial r}\left[ r^2 e^{-\nu_0/2} \Delta r \right].
\label{eq_radial_7_1}
\end{align}
%----------------------------------------------------------------------

Similarly, substituting into Eq.~\eqref{eq_radial_6_2} gives,
%----------------------------------------------------------------------
\begin{align}
(\varepsilon_0 + P_0)\frac{\partial}{\partial r}\delta\nu = \left[ \delta P - (\varepsilon_0 + P_0)\left( \frac{d\nu_0}{dr} + \frac{1}{r} \right) \Delta r \right] \frac{d}{dr}\left( \nu_0 + \lambda_0 \right).
\label{eq_radial_7_2}
\end{align}
%----------------------------------------------------------------------

At this stage, all perturbation quantities are expressed in terms of the single dynamical variable $\Delta r$, except for $\delta P$. An additional relation is therefore required, which follows from conservation laws.

\subsubsection{Conditions of Conservation}

Baryon number conservation implies $(n u^\mu)_{;\mu}=0$. Writing the baryon number density as $n(t,r)=\mathcal{N}(r)+\delta n(t,r)$, where $\mathcal{N}(r)$ is the background density, one obtains
%----------------------------------------------------------------------
\begin{align}
\delta n = - \frac{1}{\sqrt{{}^{(3)}g}} \frac{\partial}{\partial r} \left( \sqrt{{}^{(3)}g}\,\mathcal{N}\,\Delta r \right) - \frac{1}{2}\mathcal{N}\frac{\delta {}^{(3)}g}{{}^{(3)}g}.
\label{eq_cnsrv_2}
\end{align}
%----------------------------------------------------------------------

Using Eqs.~\eqref{eq_radial_3}, \eqref{eq_radial_5}, and \eqref{eq_radial_6_3}, and retaining terms up to first order, this reduces to
%----------------------------------------------------------------------
\begin{align}
\delta n = - \frac{e^{\nu_0/2}}{r^2} \frac{\partial}{\partial r} \left[ r^2 \mathcal{N} e^{-\nu_0/2} \Delta r \right].
\label{eq_cnsrv_2_1}
\end{align}
%----------------------------------------------------------------------

Assuming a cold EoS $P_0=P_0(\varepsilon_0)$, we define the adiabatic index
%----------------------------------------------------------------------
\begin{align}
\gamma =
\frac{\mathcal{N}}{P_0}\frac{dP_0}{d\mathcal{N}} = \left(1+\frac{\varepsilon_0}{P_0}\right)\frac{dP_0}{d\varepsilon_0}.
\label{eq_cnsrv_3}
\end{align}
%----------------------------------------------------------------------

Combining the above relations yields the pressure perturbation
%----------------------------------------------------------------------
\begin{align}
\delta P = -\Delta r\frac{dP_0}{dr} - 2\gamma P_0 \frac{\Delta r}{r} - \frac{\gamma P_0}{\varepsilon_0 + P_0}\frac{dP_0}{dr}\Delta r - \gamma P_0 \frac{d\Delta r}{dr}.
\label{eq_cnsrv_4}
\end{align}
%----------------------------------------------------------------------

Finally, imposing energy--momentum conservation, $T^{\mu\nu}_{~~;\nu}=0$, and assuming harmonic time dependence $e^{i\varpi t}$ for all perturbations, we obtain the radial pulsation equation
%----------------------------------------------------------------------
\begin{align}
\varpi^2 e^{\lambda_0-\nu_0}(\varepsilon_0+P_0)\Delta r &= \frac{\partial}{\partial r}\delta P + \delta P\frac{d}{dr}\left(\nu_0+\frac{\lambda_0}{2}\right) + \frac{1}{2}\delta\varepsilon\frac{d\nu_0}{dr} \nonumber\\
&\quad - \frac{1}{2}(\varepsilon_0+P_0) \left(\frac{d\nu_0}{dr}+\frac{1}{r}\right) \frac{d}{dr}(\nu_0+\lambda_0)\Delta r.
\label{eq_radial_9}
\end{align}
%----------------------------------------------------------------------

This equation constitutes a Sturm--Liouville eigenvalue problem describing radial oscillations in NSs.

\subsubsection{Pulsation Equation and Boundary Conditions}
We introduce the Lagrangian pressure perturbation $\Delta P = \delta P + \Delta r \, dP_0/dr$ and define new variables as follows \citep{Chanmugam_1977},
%----------------------------------------------------------------------
\begin{align}
\xi(r) = \frac{\Delta r}{r}, \qquad  \eta(r) = \frac{\Delta P}{P_0}.
\label{eq_bc_1}
\end{align}
%----------------------------------------------------------------------

The pulsation equations reduce to two coupled first-order equations,
%----------------------------------------------------------------------
\begin{align}
\frac{d\xi}{dr} &= -\frac{1}{r}\left(3\xi+\frac{\eta}{\gamma}\right) - \frac{\xi}{\varepsilon_0+P_0}\frac{dP_0}{dr}, \label{eq_bc_2}\\
\frac{d\eta}{dr} &= \Bigg[ \varpi^2 e^{\lambda_0-\nu_0}\frac{\varepsilon_0+P_0}{P_0}r - \frac{4}{P_0}\frac{dP_0}{dr} - 8\pi r(\varepsilon_0+P_0)e^{\lambda_0} + \frac{r}{P_0(\varepsilon_0+P_0)}\left(\frac{dP_0}{dr}\right)^2 \Bigg]\xi \nonumber\\
&\quad + \left[ -\frac{\varepsilon_0}{P_0(\varepsilon_0+P_0)}\frac{dP_0}{dr} - 4\pi r(\varepsilon_0+P_0)e^{\lambda_0} \right]\eta.
\label{eq_bc_3}
\end{align}
%----------------------------------------------------------------------

Regularity at the stellar center requires
%----------------------------------------------------------------------
\begin{align}
3\gamma\xi + \eta = 0, \qquad r \rightarrow 0,
\label{eq_bc_4}
\end{align}
while the surface boundary condition follows from $\Delta P(R)=0$,
\begin{align}
P_0(R) \,\eta(R) = 0.
\label{eq_bc_5}
\end{align}
%----------------------------------------------------------------------

The system admits a discrete spectrum of eigenvalues $\varpi_0^2 < \varpi_1^2 < \varpi_2^2 < \cdots$, with corresponding eigen functions $\xi_0, \xi_1, \xi_2, \ldots$. A stellar configuration is stable against radial perturbations if $\varpi_n^2 > 0$ for all $n$, whereas $\varpi_0^2 < 0$ signals the onset of dynamical instability.

\section{Numerical Results and Discussion}
\label{sec:3}

We employ two distinct frameworks, namely NL-RMF and DD-RMF, to construct purely nucleonic NSs composed of neutrons, protons, electrons, and muons. Within the NL-RMF approach, we adopt the NL3, GMT, and IUFSU parameterizations, while the DD-RMF description includes the DD2, PKDD, and DDME$\delta$ parameterizations. In addition to nucleons, antikaons ($K^-$ and $\bar{K}^0$) are incorporated as additional degrees of freedom, allowing for the possible formation of kaon-condensed phases in dense matter. The antikaon optical potential is varied over the range $U_{\bar{K}} = -120$ to $-160$ MeV in order to assess its influence on the stellar composition and dynamical properties.

The nature of the phase transition to antikaon-condensed matter depends sensitively on how rapidly the in-medium antikaon energy decreases relative to the electron chemical potentials. In the case of a second-order transition, the $K^-$ condensate forms continuously beyond the threshold density, resulting in a smooth softening of the EoS. In contrast, a first-order transition is characterized by the abrupt onset of the condensate, accompanied by a discontinuous change in composition and a more pronounced reduction in pressure. Our calculations indicate that the transition to the antikaon-condensed phase is of second order for all EoSs considered, with the exception of the DDME$\delta$ parametrization. For this model, the transition becomes first order for antikaon optical potentials in the range $U_{\bar{K}} = -130$ to $-160$ MeV.

When a first-order transition occurs, we determine the phase boundary using the Maxwell construction, imposing equality of pressure and baryon chemical potential between the purely nucleonic and kaon-condensed phases. This construction introduces a density discontinuity across the transition, further enhancing the softening of the EoS and potentially leaving discernible imprints on NS structure and oscillation properties.  Finally, for all models and values of the antikaon optical potential considered, we find that the negatively charged antikaon $K^-$ condenses prior to the neutral $\bar{K}^0$. This behavior follows from the charge-neutrality and $\beta$-equilibrium conditions: the rising electron chemical potential lowers the effective threshold for $K^-$ condensation, while the neutral $\bar{K}^0$, which does not couple to the charge chemical potential, requires higher densities to become energetically favored.
%----------------------------------------------------------------------
\begin{figure}[ht]
	\includegraphics[width=\linewidth]{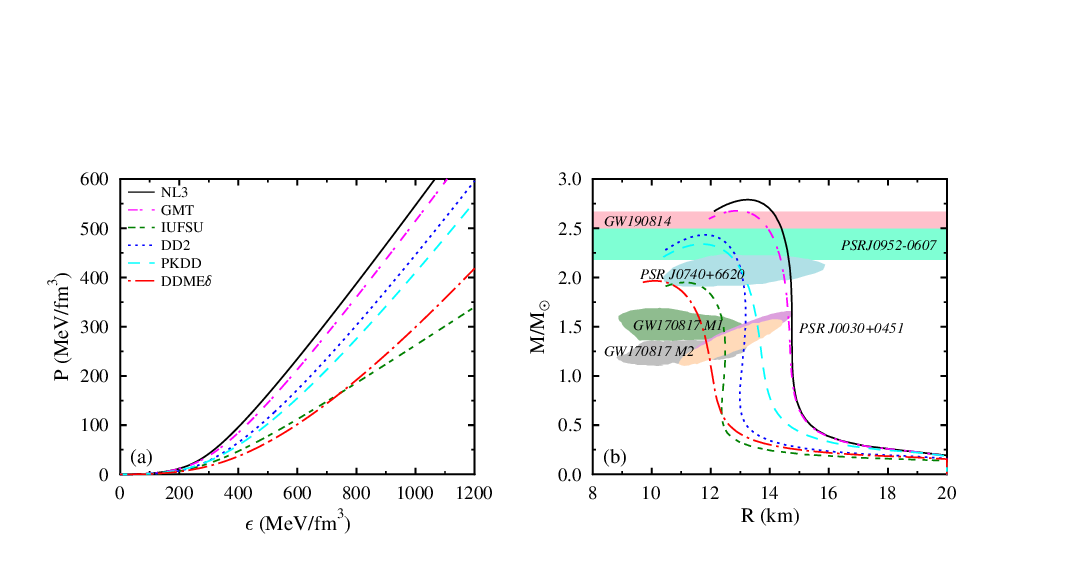}
	\caption{(a) Pressure as a function of energy density for matter composed exclusively of nucleons ($n,p$) and leptons ($e^-,\mu^-$). (b) Corresponding mass--radius relations derived from the EoSs obtained using the NL3, GMT, IUFSU, DD2, PKDD, and DDME$\delta$ parametrizations. 
    %Observational constraints on NS masses and radii from PSR J0740+6620 \citep{Fonseca:2021wxt,Miller:2021qha}, PSR J0030+0451 \citep{Riley:2019yda,Miller:2019cac}, and PSR J0952$-$0607 \citep{Romani:2022jhd} are indicated. The shaded regions on the mass--radius plane correspond to the constraints inferred from the gravitational-wave events GW170817 \citep{LIGOScientific:2018cki} and GW190814 \citep{LIGOScientific:2020zkf}.
    }
	\label{EoS_MR_RMF}
\end{figure}  
%----------------------------------------------------------------------

Figure~\ref{EoS_MR_RMF} displays the pressure--energy density relations and the corresponding mass--radius sequences for NS matter composed of nucleons and leptons ($n$, $p$, $e^-$, $\mu^-$), in the absence of antikaon condensation, obtained using the RMF parametrizations considered in this work. Both NL and DD-RMF models are employed, enabling a systematic exploration of EoSs with differing stiffness at supranuclear densities. The resulting mass--radius relations are confronted with current observational constraints from precision pulsar measurements, including PSR J0740$+$6620 \citep{Fonseca:2021wxt, Miller:2021qha}, PSR J0030$+$0451 \citep{Riley:2019yda, Miller:2019cac}, and PSR J0952$-$0607 \citep{Romani:2022jhd}, as well as bounds inferred from the gravitational-wave events GW170817 \citep{LIGOScientific:2018cki} and GW190814 \citep{LIGOScientific:2020zkf}. 

The NL-RMF parametrizations NL3 and GMT, together with the DD-RMF models DD2 and PKDD, produce comparatively stiffer EoSs, resulting in larger predicted NS radii due to enhanced pressure support at high densities. Consequently, these models are more consistent with high-mass observational constraints, such as those from PSR J0952$-$0607, PSR J0740$+$6620, and GW190814, but are less compatible with constraints favoring smaller radii at lower masses. In contrast, the IUFSU and DDME$\delta$ parametrizations yield softer EoSs, leading to more compact stellar configurations with reduced radii. These models are better aligned with low-mass constraints, such as those inferred from GW170817 and PSR J0030$+$0451, but fail to simultaneously satisfy the highest observed NS mass limits. Overall, this comparison highlights how variations in the high-density behavior of the EoS manifest in NS structure and helps identify models that are consistent with current astrophysical observations.

\begin{figure}[ht]
	\includegraphics[width=\linewidth]{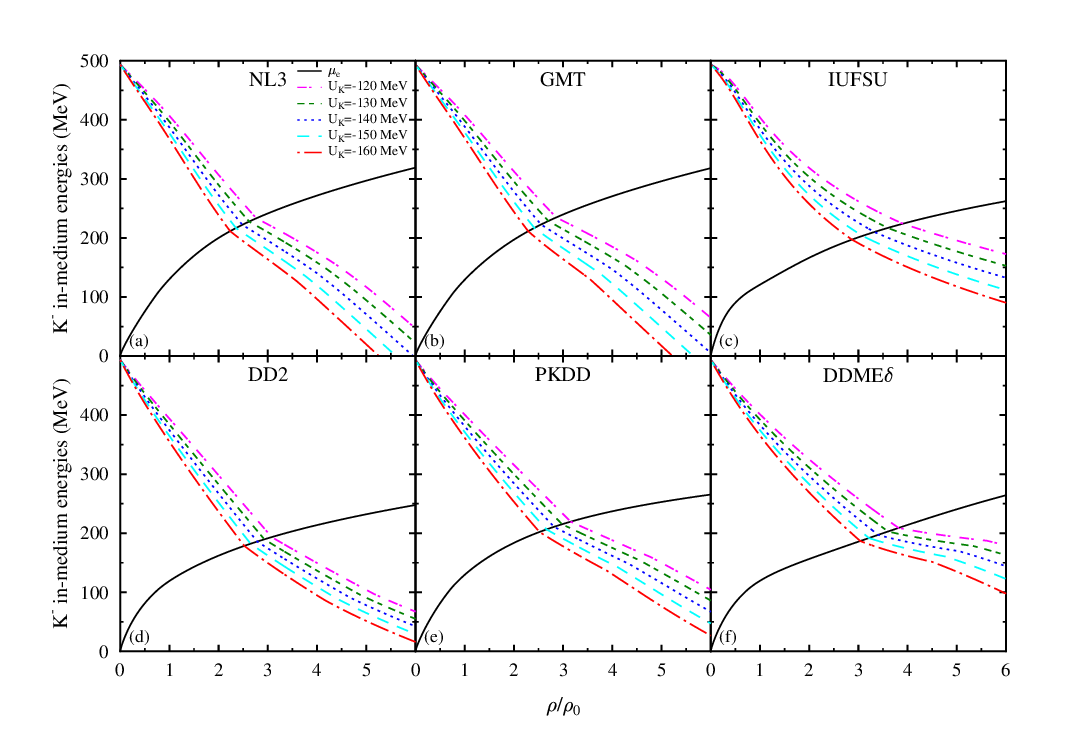}
	\caption{In-medium energies of the $K^-$ meson as a function of baryon density $\rho_B$ (normalized to the nuclear saturation density $\rho_0$) for different RMF parametrizations, shown for several values of the antikaon optical potential depth $U_{\bar{K}}$.}
	\label{K-energy_RMF_kaon}
\end{figure}

\begin{figure}[ht]
	\includegraphics[width=\linewidth]{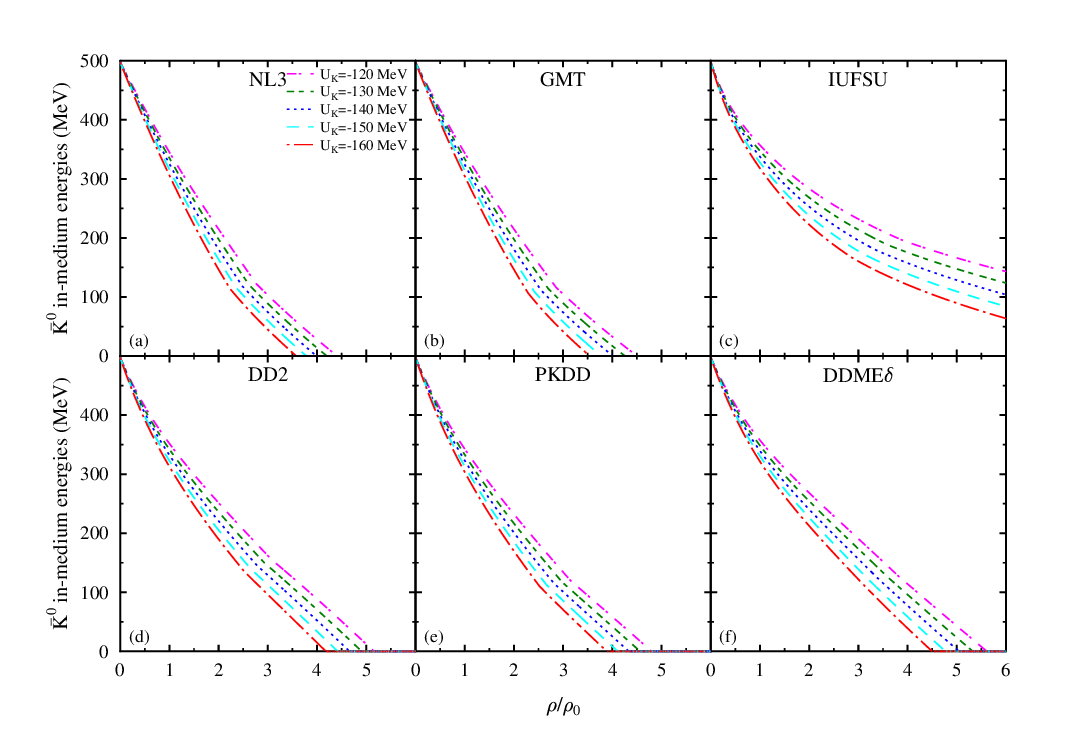}
	\caption{The in-medium energies of $\bar{K^0}$ as functions of baryon density $\rho_B$ (normalized to the nuclear saturation density $\rho_0$) for various RMF parametrizations, shown for different choices of the antikaon optical potential depth $U_{\bar{K}}$.}
	\label{K0bar_energy_RMF_kaon}
\end{figure}

Figures~\ref{K-energy_RMF_kaon} and \ref{K0bar_energy_RMF_kaon} show the density dependence of the in-medium energies of $K^-$ and $\bar{K}^0$ mesons, respectively, for antikaon optical potentials in the range $U_{\bar{K}}=-120$ to $-160$ MeV and for the NL3, GMT, IUFSU, DD2, PKDD, and DDME$\delta$ parameter sets. For all models, the effective energies of both antikaons decrease monotonically with increasing baryon density, although the rate of decrease and the corresponding condensation thresholds exhibit notable model dependence.

The onset of $K^-$ condensation is determined by the crossing of the in-medium energy $\omega_{K^-}$ with the electron chemical potential $\mu_e$, marking the transition from a purely nucleonic phase to a kaon-condensed phase. In contrast, the $\bar{K}^0$ condensate forms when in-medium energy $\omega_{\bar{K}^0}$ vanishes, which occurs at systematically higher densities than the $K^-$ threshold for all EoSs. Increasing the magnitude of the attractive antikaon optical potential lowers the in-medium energies and shifts both condensation thresholds to lower densities across all models.

%----------------------------------------------------------------------
%----------------------------------------------------------------------
%----------------------------------------------------------------------
%----------------------------------------------------------------------

The NL3 and GMT parametrizations exhibit larger electron chemical potentials at moderate baryon densities, a direct consequence of their stiffer symmetry energy. Consequently, for a fixed value of $U_{\bar{K}}$, the crossing between $\mu_e$ and the in-medium $K^-$ energy occurs at lower densities, leading to an earlier onset of $K^-$ condensation. The PKDD model shows a similar behavior, although the $\bar{K}^0$ condensate sets in at comparatively lower densities as compared to DD2 and DDME$\delta$ due to higher symmetric energy. In contrast, the IUFSU and DDME$\delta$ parametrizations yield smaller electron chemical potentials, thereby shifting the onset of $K^-$ condensation to higher baryon densities. The appearance of $\bar{K}^0$ condensates is likewise delayed in these models. The threshold densities for antikaon condensation across different EoSs and optical potentials are summarized in Table~\ref{tab:6}, illustrating a systematic decrease in the onset densities with increasing strength of the antikaon optical potential at saturation density $\rho_0$.
Our results are consistent with previous studies of antikaon condensation such as in-medium energies, threshold densities  and particle compositions \citep{Banik:2000dx, Banik:2001yw, Banik:2002qu, Gupta:2013rba, Char:2014cja, Batra:2017mfv} in NSs. In all models, the in-medium energies of $K^-$ and $\bar{K}^0$ decrease with increasing density, leading to the appearance of $K^-$ prior to $\bar{K}^0$, while deeper optical potentials shift both condensation thresholds to lower densities. Models with stiffer symmetry energy (NL3, GMT, PKDD) predict earlier condensation through larger values of $\mu_e$, whereas softer parameterizations (IUFSU, DDME$\delta$) delay the onset.

%----------------------------------------------------------------------
\begin{figure}[ht]
	\includegraphics[width=\linewidth]{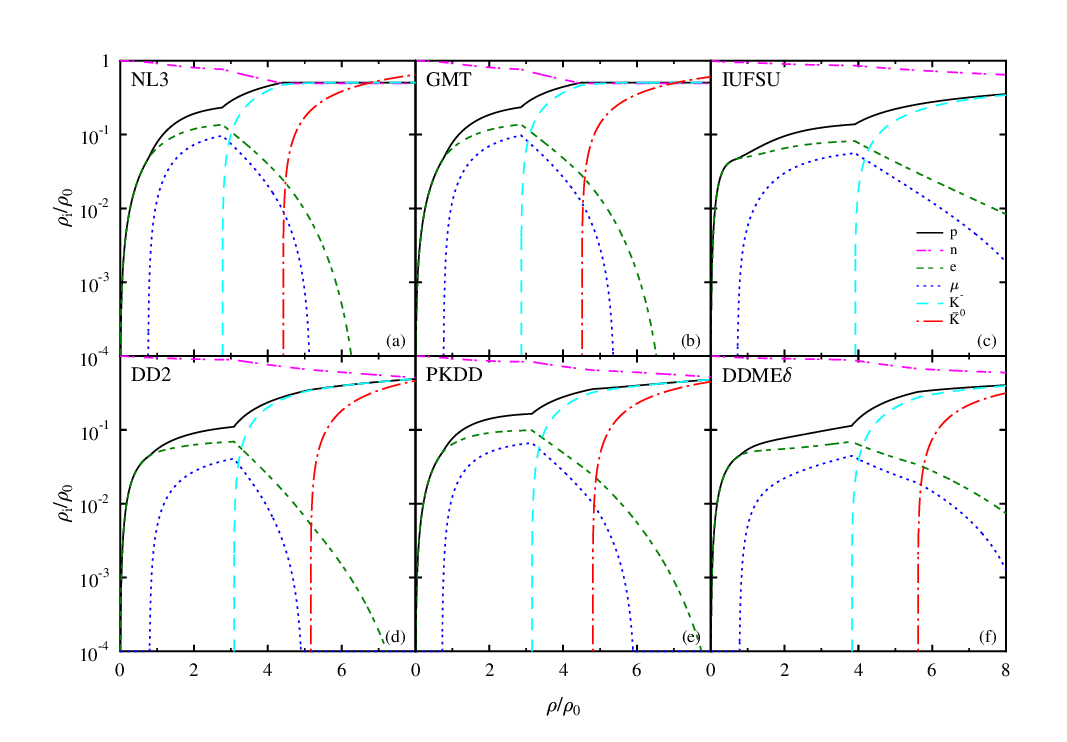}
	\caption{Population fractions $\rho_i/\rho_0$ of nucleons, leptons, and antikaons as functions of the baryon density $\rho_B/\rho_0$, calculated using different RMF parametrizations for an antikaon optical potential depth of $U_{\bar{K}} = -120$ MeV.}
	\label{particle_frac_120}
\end{figure}
%----------------------------------------------------------------------

The particle composition of NS matter in the presence of antikaons is shown in figure~\ref{particle_frac_120} as a function of baryon density for an antikaon optical potential of $U_{\bar{K}} = -120$ MeV, across all EoS families presented in figure~\ref{EoS_MR_RMF}. At low baryon densities, the matter is dominated by nucleons and leptons, with charge neutrality maintained by protons, electrons, and muons. When the baryon density exceeds the corresponding threshold density for antikaon condensation, antikaons start to appear, resulting in a substantial reorganization of the matter composition. In particular, the onset of $K^-$ condensation causes a rapid depletion of the lepton fractions, as negatively charged antikaons effectively replace electrons and muons in maintaining charge neutrality.

The population profiles predicted by the NL3, GMT, and PKDD parametrizations are qualitatively similar, differing mainly in the onset densities of the $\bar{K}^0$ condensate. Owing to their stiffer symmetry energy, these models predict larger proton fractions at lower densities, which in turn lead to enhanced electron and muon populations compared to the IUFSU, DD2, and DDME$\delta$ parametrizations. 

We also observe model dependence in the particle composition and it is more pronounced at higher densities.  For instance, in the PKDD parametrization, $\bar{K}^0$ mesons appear at comparatively lower densities than in the DD2 model, while the suppression of leptons proceeds more gradually. This prolonged presence of leptons, together with the early onset of antikaons, contributes to an enhanced softening of the EoS. For a fixed value of $U_{\bar{K}}$, the onset of $K^-$ condensation is shifted to higher baryon densities in the IUFSU and DDME$\delta$ models. In IUFSU, this delay arises from the softened density dependence of the symmetry energy produced by the isoscalar--isovector coupling $\Lambda_{\rm v}$, which reduces the electron chemical potential and therefore raises the threshold for $K^-$ condensation. In DDME$\delta$, the inclusion of the $\delta$-meson channel modifies the neutron--proton effective mass splitting and weakens the growth of the isospin asymmetry, leading again to smaller electron chemical potentials and a delayed onset of antikaon condensation.

%----------------------------------------------------------------------
\begin{table}[ht!] 
\centering
\caption{Threshold densities, $\rho_{\rm{th}}$ (in units of $\rho_0$), corresponding to the onset of antikaon condensation in NS matter for various values of $U_{\bar{K}}$, for NL and DD models.}
	\begin{tabular}{cccccccc}
		\hline \hline
		$U_{\bar{K}}$ (MeV) & {} & NL3 & GMT & IUFSU & DD2 & PKDD & DDME$\delta$ \\
		\hline
		& $\rho_{\rm th}(K^-)$ & 2.771 & 2.863 & 3.910 & 3.080 & 3.157 & 3.823 \\
		\shortstack{-120\\~} & $\rho_{\rm th}(\bar{K}^0)$ & 4.413 & 4.511 & -- & 5.160 & 4.802 & 5.612 \\
		\hline
		& $\rho_{\rm th}(K^-)$ & 2.622 & 2.704 & 3.601 & 2.932 & 2.983 & 3.606 \\
		\shortstack{-130\\~} & $\rho_{\rm th}(\bar{K}^0)$ & 4.203 & 4.263 & -- & 4.911 & 4.568 & 5.329 \\
		\hline
		& $\rho_{\rm th}(K^-)$ & 2.487 & 2.560 & 3.330 & 2.791 & 2.822 & 3.402 \\
		\shortstack{-140\\~} & $\rho_{\rm th}(\bar{K}^0)$ & 3.994 & 4.021 & -- & 4.663 & 4.340 & 5.047 \\
		\hline
		& $\rho_{\rm th}(K^-)$ & 2.358 & 2.421 & 3.084 & 2.657 & 2.669 & 3.218 \\
		\shortstack{-150\\~} & $\rho_{\rm th}(\bar{K}^0)$ & 3.784 & 3.780 & -- & 4.421 & 4.113 & 4.770 \\
		\hline
		& $\rho_{\rm th}(K^-)$ & 2.237 & 2.297 & 2.872 & 2.523 & 2.521 & 3.040 \\
		\shortstack{-160\\~} & $\rho_{\rm th}(\bar{K}^0)$ & 3.581 & 3.545 & -- & 4.180 & 3.892 & 4.501 \\
		\hline \hline
	\end{tabular}
	\label{tab:6}
\end{table}
%----------------------------------------------------------------------

%----------------------------------------------------------------------
\begin{figure}[ht]
	\includegraphics[width=\linewidth]{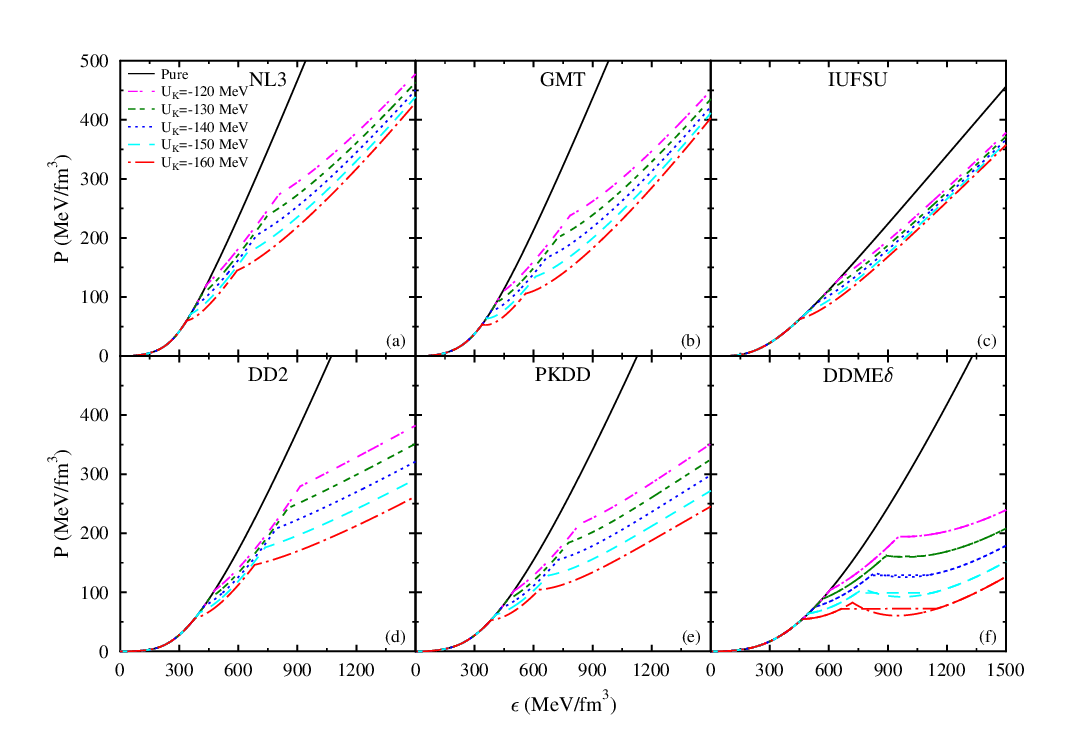}
	\caption{Pressure as a function of energy density (EoS) for various RMF parametrizations, illustrated for different choices of the antikaon optical potential depth $U_{\bar{K}}$.}
	\label{EoS_RMF_kaon}
\end{figure} 
%----------------------------------------------------------------------
Figure~\ref{EoS_RMF_kaon} shows the pressure as a function of energy density for NS matter constructed within both the NL and DD RMF frameworks. In the absence of antikaon condensation, the NL3 parametrization yields the stiffest EoS, reflecting its strong repulsive vector interactions at high densities, whereas the IUFSU parametrization produces the softest EoS due to its enhanced isovector and nonlinear couplings, which reduce the pressure at supranuclear densities.

The inclusion of antikaons systematically softens the EoS, with the effect becoming more pronounced as the attractive strength of the $K^-$ optical potential increases. This softening is manifested by distinct changes in the slope of the pressure--energy density relation: the first kink marks the onset of $K^-$ condensation, while a second kink--when present--signals the appearance of $\bar{K}^0$. While the $K^-$ threshold occurs for all parametrizations considered, the $\bar{K}^0$ condensate does not emerge in the IUFSU model within the explored density range.

Within the NL-RMF models, the IUFSU parametrization predicts the onset of $K^-$ condensation at higher densities compared to the NL3 and GMT parametrizations for a fixed value of $U_{\bar{K}}$, consistent with its softer symmetry energy. In the DD-RMF framework, DD2 produces the stiffest EoS, whereas DDME$\delta$ remains the softest. This behavior reflects differences in the density dependence of the meson--nucleon couplings, as well as the role of the $\delta$ meson in enhancing isospin asymmetry effects at high densities.

%----------------------------------------------------------------------
\begin{figure}[ht]
	\includegraphics[width=\linewidth]{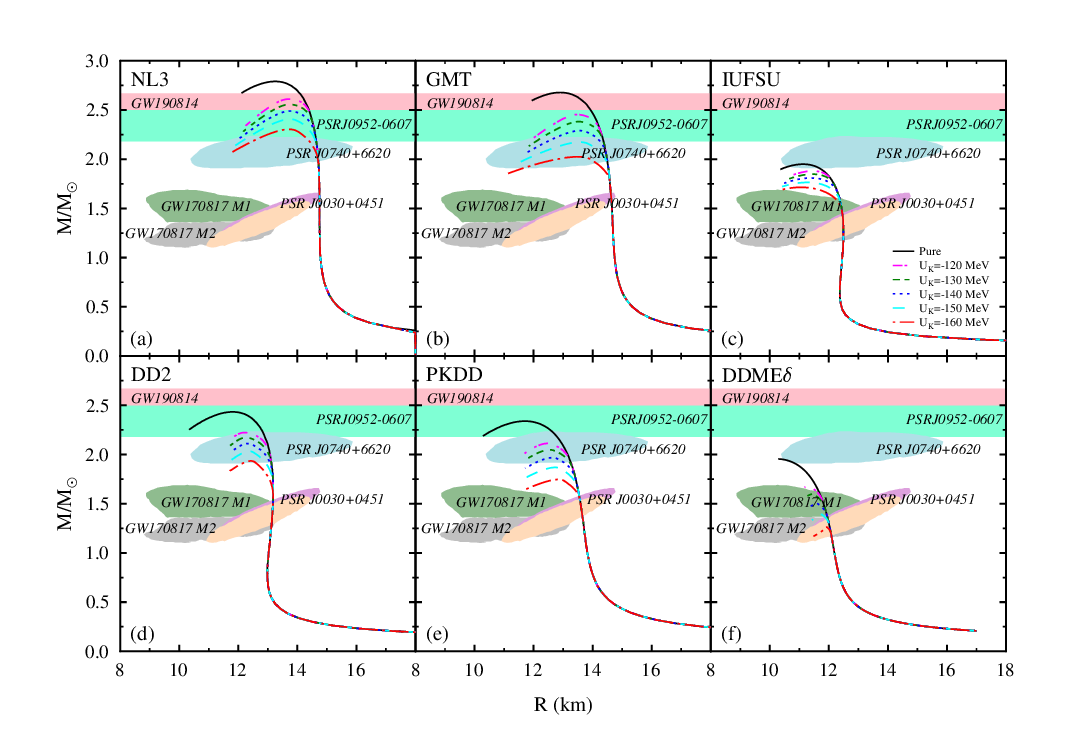}
	\caption{Mass--radius relations for different RMF parametrizations and antikaon optical potential depths $U_{\bar{K}}$, compared with observational constraints from PSR J0740+6620 \citep{Fonseca:2021wxt,Miller:2021qha}, PSR J0030+0451 \citep{Riley:2019yda,Miller:2019cac}, PSR J0952--0607 \citep{Romani:2022jhd}, and the GW170817 \citep{LIGOScientific:2018cki} and GW190814 \citep{LIGOScientific:2020zkf} events.}
	\label{MR_RMF_kaon}
\end{figure}
%----------------------------------------------------------------------
Figure~\ref{MR_RMF_kaon} presents the mass--radius relations of static, spherically symmetric NSs obtained by solving the TOV equations using the EoSs shown in figure~\ref{EoS_RMF_kaon}. For the neutron-star crust, we adopt the Haensel--Pichon EoS for the outer crust \citep{Haensel:1993zw} and the SLy EoS for the inner crust \citep{Douchin:2001sv}, which provides a unified and thermodynamically consistent description of the crust--core interface. The corresponding maximum masses and radii for different values of the antikaon optical potential depth, $U_{\bar{K}}$, are summarized in table~\ref{tab:7} for both purely nucleonic matter and matter containing antikaons.

For matter composed solely of nucleons and leptons, the predicted maximum masses are $2.790~M_{\odot}$, $2.677~M_{\odot}$, and $1.949~M_{\odot}$ for the NL-RMF parametrizations NL3, GMT, and IUFSU, respectively. In the DD-RMF framework, the DD2, PKDD, and DDME$\delta$ parametrizations yield maximum masses of $2.432~M_{\odot}$, $2.338~M_{\odot}$, and $1.954~M_{\odot}$. These differences directly reflect the varying stiffness of the underlying EoSs at high baryon densities.

The inclusion of antikaon condensation leads to a substantial reduction in the maximum mass across all parametrizations. This reduction becomes increasingly pronounced for more attractive antikaon optical potentials, as deeper potentials trigger an earlier onset of condensation and enhance the softening of the EoS at supranuclear densities. At the canonical mass of $1.4~M_{\odot}$, stiffer EoSs such as NL3 predict larger stellar radii ($\gtrsim 13.5$ km), whereas softer parametrizations, such as IUFSU and DDME$\delta$, yield more compact configurations with radii in the range $\sim$12--13 km. This behavior improves consistency with NICER radius measurements \citep{Riley:2019yda, Miller:2019cac, Miller:2021qha} and gravitational-wave constraints from GW170817 \citep{LIGOScientific:2018cki}.

However, sufficiently deep antikaon optical potentials can lead to an enhanced softening of the EoS, which may lower the maximum NS mass below the observational lower bound of $\sim 2~M_{\odot}$ inferred from massive pulsars \citep{Fonseca_2021_SOyMc, Romani:2022jhd}. Within this context, the $2~M_{\odot}$ constraint is satisfied in the DD2 parametrization up to $U_{\bar{K}}=-150$ MeV and in PKDD up to $U_{\bar{K}}=-130$ MeV. In contrast, the NL3 and GMT parametrizations preserve compatibility with the maximum-mass constraint over a broader range of antikaon optical potentials, owing to their intrinsically stiffer behavior at high densities.

Overall, antikaon condensation substantially modifies the high-density behavior of dense matter, with direct consequences for NS structure, most notably the maximum mass and stellar radius. When combined with the observational limits, these structural effects impose stringent constraints on the permissible strength of antikaon interactions in NS interiors.
%----------------------------------------------------------------------
\begin{table}[ht] 
\centering
\caption{Maximum mass, $M_{\rm max}$ (in units of $(M_\odot)$), corresponding radius (in km) of NS for different value of $K^-$ optical potential depths $U_{\bar{K}}$ (in units of MeV), for NL and  RMF models.}
\begin{tabular}{cccccccc}
\hline \hline
$U_{\bar{K}}$ (MeV) & {} & NL3 & GMT & IUFSU & DD2 & PKDD & DDME$\delta$ \\
\hline
  &
$M_{\rm max}(M_\odot)$ & 2.790 & 2.677 & 1.949 & 2.432 & 2.338 & 1.954 \\
\shortstack{0\\~} & $R$ (km) & 13.23 & 12.91 & 11.12 & 11.79 & 11.71 & 10.29 \\
\hline
 &
$M_{\rm max}(M_\odot)$ & 2.610 & 2.454 & 1.880 & 2.223 & 2.112 & 1.676 \\
\shortstack{-120\\~}  & $R$ (km) & 13.74 & 13.48 & 11.46 & 12.21 & 12.45 & 11.31 \\
\hline
 & $M_{\rm max}(M_\odot)$ 
           & 2.557 & 2.382 & 1.848 & 2.173 & 2.047 & 1.598 \\
\shortstack{-130\\~} & $R$ (km) & 13.73 & 13.53 & 11.44 & 12.27 & 12.55 & 11.45 \\
\hline
 & $M_{\rm max}(M_\odot)$ & 2.490 & 2.290 & 1.809 & 2.110 & 1.967 & 1.503 \\
\shortstack{-140\\~} & $R$ (km) & 13.76 & 13.57 & 11.40 & 12.32 & 12.64 & 11.60 \\
\hline
 & $M_{\rm max}(M_\odot)$ & 2.407 & 2.173 & 1.764  & 2.032& 1.868 & 1.391 \\
\shortstack{-150\\~} & $R$ (km) & 13.75 & 13.58 & 11.26 & 12.38 & 12.72 & 11.76 \\
\hline
 & $M_{\rm max}(M_\odot)$ & 2.304 & 2.024 & 1.714 & 1.934 & 1.746 & 1.263 \\
\shortstack{-160\\~}& $R$ (km) & 13.72 & 13.60 & 11.07 & 12.42 & 12.83 & 11.93 \\
\hline \hline
\end{tabular}
\label{tab:7}
\end{table}
%----------------------------------------------------------------------

%----------------------------------------------------------------------
\begin{figure}[ht]
	\includegraphics[width=\linewidth]{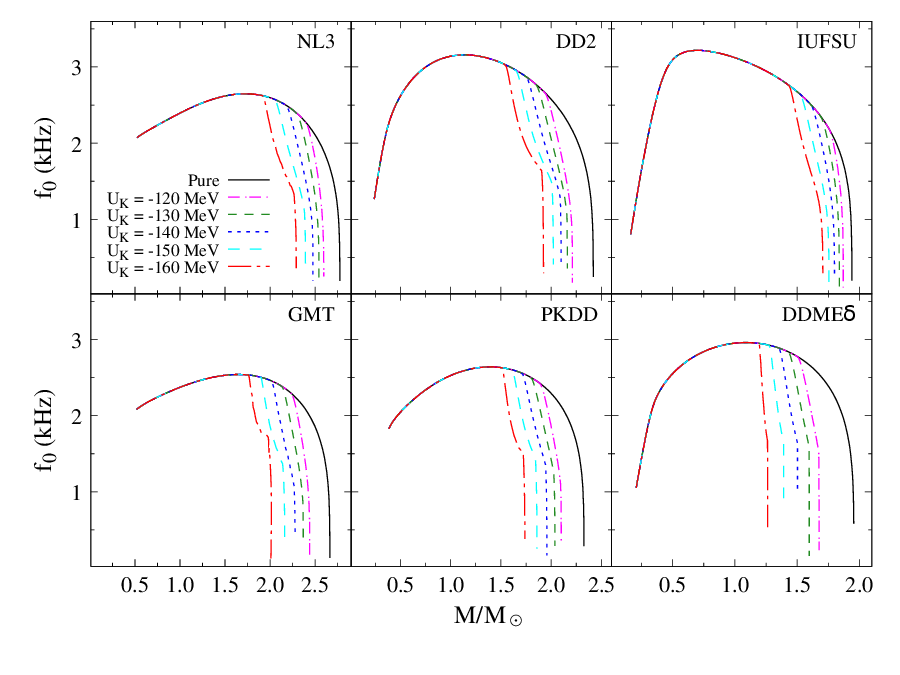}
	\caption{Plot of fundamental radial mode frequencies, f$_0$, against NS masses. The solid black line shows the results for purely nucleonic NSs. The other curves show the results for NS models with antikaon condensates, assuming various $U_{\bar{K}}$ values.}
	\label{f0_vs_M}
\end{figure}
%----------------------------------------------------------------------

%----------------------------------------------------------------------
\begin{figure}[ht]
	\includegraphics[width=\linewidth]{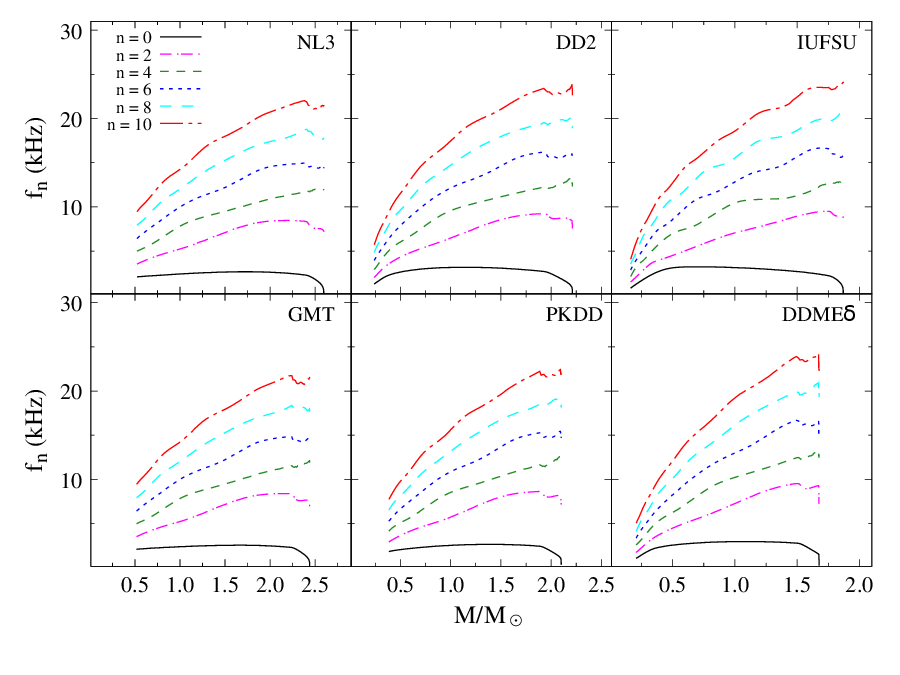}
	\caption{Frequencies of fundamental radial mode (solid black line) and other higher order modes with respect to NS masses. The results shown for NS models are obtained taking $U_{\bar{K}}=-120$ MeV.}
	\label{f_vs_M}
\end{figure}
%----------------------------------------------------------------------
%----------------------------------------------------------------------
\begin{figure}[ht!]
	\includegraphics[width=\linewidth]{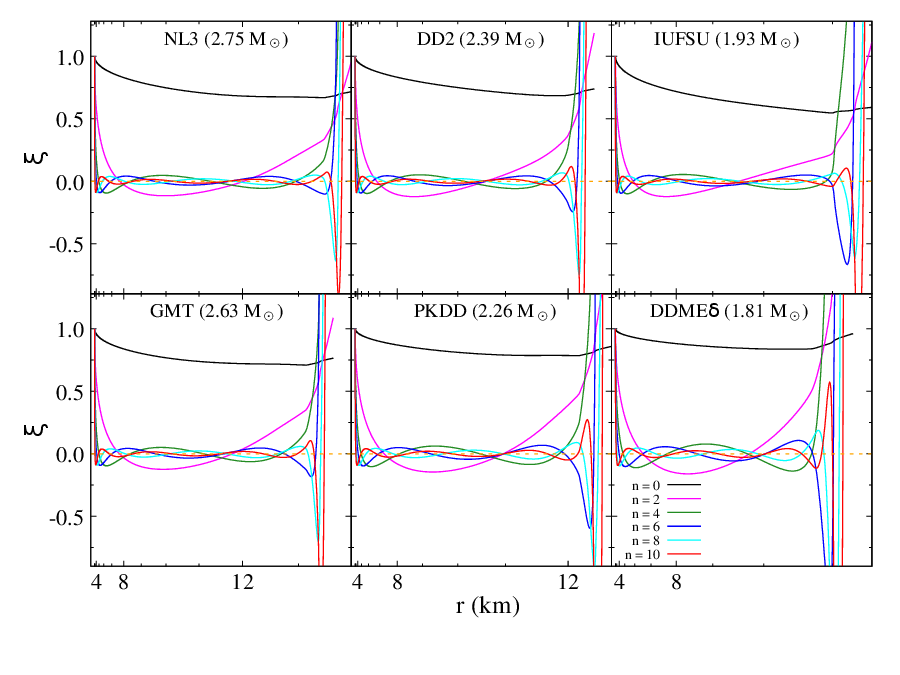}
	\caption{Radial displacement eigenfunction $\xi(r)$ for purely nucleonic NS models at the indicated stellar masses, shown for the fundamental radial mode (solid black line) and selected higher-order modes. The displacement is normalized such that $\xi(0)=1$ for all modes. The horizontal dotted line denotes the zero-displacement level.}
	\label{chi_NS}
\end{figure}
%----------------------------------------------------------------------
%----------------------------------------------------------------------
\begin{figure}[ht!]
	\includegraphics[width=\linewidth]{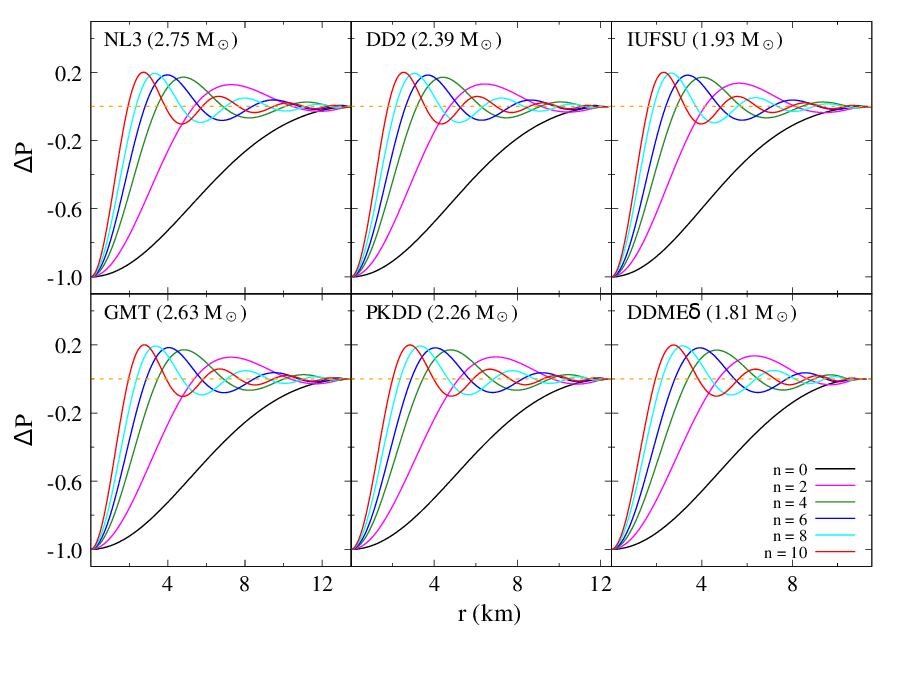}
	\caption{Lagrangian pressure perturbation $\Delta P(r)$ for purely nucleonic NS models, shown for the fundamental radial mode (solid black line) and selected higher-order modes. The perturbation is rendered dimensionless and normalized such that $\Delta P = -1$ at the stellar center. The horizontal dotted line indicates the zero-perturbation level.}
	\label{delp_NS}
\end{figure}
%----------------------------------------------------------------------

%----------------------------------------------------------------------
\begin{figure}[ht!]
	\includegraphics[width=\linewidth]{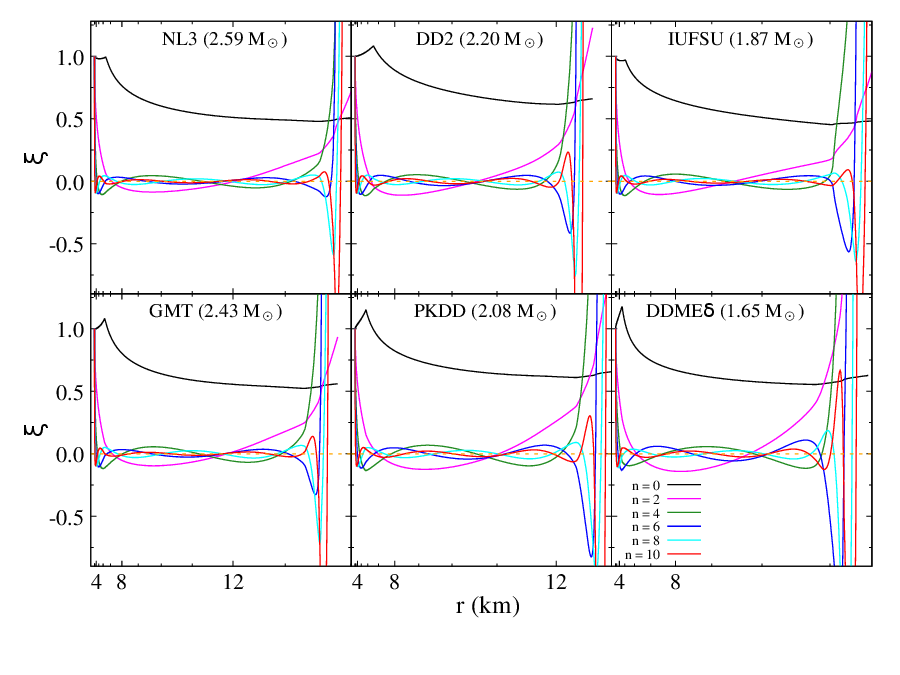}
	\caption{Radial displacement $\xi(r)$ for NS models containing antikaon condensates, shown for the fundamental radial mode (solid black line) and selected higher-order modes. The displacement is normalized such that $\xi(0)=1$ at the stellar center. All configurations correspond to an antikaon optical potential depth of $U_{\bar{K}}=-120$ MeV. The horizontal dotted line indicates the zero-displacement level.}
	\label{chi_antikaon}
\end{figure}
%----------------------------------------------------------------------
%----------------------------------------------------------------------
\begin{figure}[ht!]
	\includegraphics[width=\linewidth]{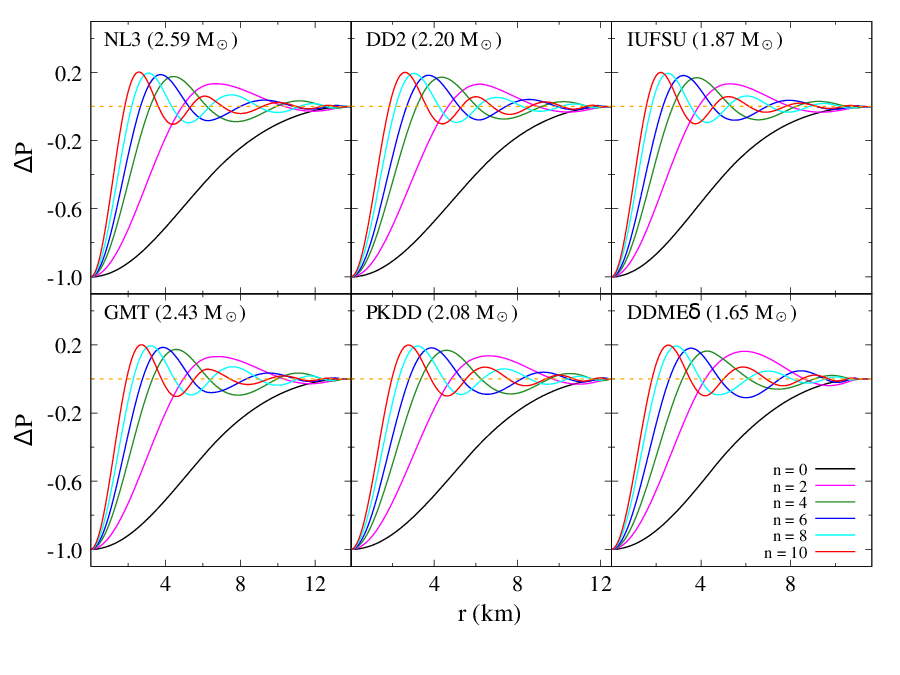}
	\caption{Lagrangian pressure perturbation $\Delta P(r)$ for NS models containing antikaon condensates, shown for the fundamental radial mode (solid black line) and selected higher-order modes. The perturbation is rendered dimensionless and normalized such that $\Delta P(0) = -1$. All configurations correspond to an antikaon optical potential depth of $U_{\bar{K}} = -120$ MeV. The horizontal dotted line indicates the zero-perturbation level.}
	\label{delp_antikaon}
\end{figure}
%----------------------------------------------------------------------

In figure~\ref{f0_vs_M}, we present the fundamental radial oscillation frequencies ($\mathrm{f}_0 = \varpi_0/2\pi$) of NSs as a function of stellar mass. Numerical solutions are obtained for NS models constructed using the EoSs discussed in sections~\ref{sec:2_A} and \ref{sec:2_B}. For each RMF parametrization, we consider both purely nucleonic matter and matter containing antikaons, spanning a range of antikaon optical potential depths listed in table~\ref{tab:5}. The equations governing radial pulsations are solved for all stellar configurations, and the resulting frequencies are compared between nucleonic NSs and those containing antikaon condensates.

For purely nucleonic NSs, the fundamental radial oscillation frequency $\mathrm{f}0$ initially increases with stellar mass. For stiff EoSs such as NL3 and GMT, this rising trend persists up to stellar masses of approximately $1.75~M_{\odot}$. In case of DD2 and DDME$\delta$ EoSs, the maximum of $\mathrm{f}_0$ is reached at around $1.25~M_{\odot}$, while for PKDD and IUFSU the corresponding values are approximately $1.5~M_{\odot}$ and $0.6~M_{\odot}$, respectively. Beyond these masses, $\mathrm{f}_0$ decreases monotonically with increasing stellar mass for all EoS models, approaching zero for the maximally stable NS configuration against radial perturbations.

The $K^{-}$ condensate appears in the medium when its in-medium energy crosses the electron chemical potential. As shown in figure~\ref{K-energy_RMF_kaon}, increasing the magnitude of the antikaon optical potential facilitates an earlier onset of $K^{-}$ condensation. The presence of $K^{-}$ mesons in the NS core softens the EoS, leading to increased stellar compactness across all EoS families (see figure~\ref{MR_RMF_kaon}). This softening is directly reflected in the behavior of the fundamental oscillation frequency: once $K^{-}$ condensation sets in, $\mathrm{f}_0$ decreases more rapidly with increasing stellar mass and vanishes at a lower mass compared to the purely nucleonic case. This suppression of oscillation frequencies is particularly pronounced for the DDME$\delta$ parametrization. In this model, the coupling between the $\delta$ meson and antikaons significantly modifies the effective masses and elevates the in-medium antikaon energies, thereby delaying the onset of condensation to higher densities. When antikaons eventually appear, the associated reduction in pressure is substantial, leading to a marked softening of the EoS at high densities. Consequently, for NSs of comparable mass, a systematically lower fundamental oscillation frequency may be consistently interpreted as a signature of an antikaon-condensed phase in the stellar interior.

The subsequent onset of the $\bar{K}^0$ condensate leads to further softening of the EoS and manifests as an additional change in the slope of the $\mathrm{f}_0$--$M$ relation. This effect becomes increasingly prominent with increasing optical potential depth; for $U_{\bar{K}} = -160$ MeV, the change in slope is particularly sharp. In NS models based on the DDME$\delta$ parametrization, the appearance of $\bar{K}^0$ produces a comparatively mild additional change in the slope, even for the deepest optical potentials considered. 

The fundamental oscillation frequency vanishes at the maximum-mass configuration, $M_{\mathrm{max}}$, signaling the onset of instability against radial perturbations, mimicking the rapid transition results of ref. \citep{Pereira:2018}. We find that $M_{\mathrm{max}}$ decreases systematically with the appearance of antikaon condensates and with increasing optical potential depth. Among the EoSs considered, NL3 yields the largest $M_{\mathrm{max}}$ in the absence of antikaons, with the maximum mass decreasing progressively as the optical potential depth increases. For purely nucleonic matter, the IUFSU and DDME$\delta$ EoSs yield $M_{\mathrm{max}} \simeq 2~M_\odot$. The inclusion of antikaon condensates reduces $M_{\mathrm{max}}$ in these models as well; notably, for the DDME$\delta$ EoS, the maximum mass decreases to $\simeq 1.25~M_\odot$ for the largest optical potential depth considered.

In figure~\ref{f_vs_M}, we present the oscillation frequencies of higher-order radial modes ($\mathrm{f}_n = \varpi_n/2\pi$) as functions of stellar mass. We show modes up to $n = 10$ for NS models containing antikaons in their cores, fixing the antikaon optical potential depth at $U_{\bar{K}} = -120$ MeV. The qualitative behavior remains unchanged for other choices of the optical potential. As in the case of the fundamental mode, the onset of antikaon condensation induces noticeable changes in the slopes of the $\mathrm{f}_n$--$M$ relations. Moreover, higher-order modes exhibit a rapid decrease in frequency as the stellar mass approaches $M_{\mathrm{max}}$, corresponding to the maximum mass of radially stable NS configurations, as determined by the vanishing of the fundamental ocillation mode.

The eigenfunctions associated with the $n^{\text{th}}$ radial mode possess $n$ nodes between the stellar center and the surface, whereas the fundamental mode has no internal nodes. In figure~\ref{chi_NS}, we show the radial profiles of the displacement function $\xi(r)$ for modes up to $n = 10$ for purely nucleonic NS configurations, considering a few representative stellar masses. For the chosen models, the amplitude of the radial displacement $\xi(r)$ decreases monotonically with radius. Near the stellar surface, where the crust appears, the adiabatic index decreases sharply, leading to a discontinuity in the right-hand side of equation~\ref{eq_bc_2}. As a consequence, we observe a slight change in the slope of the $\xi(r)$ profiles in the vicinity of the core--crust transition.

For higher-order modes, the displacement function $\xi(r)$ crosses zero multiple times within the crust, indicating very short radial wavelengths in this region. This behavior points to the presence of high-frequency crustal oscillations and suggests a possible separation of oscillations as core modes and surface modes \citep{DiClemente2020, Gondek:1999ad}. The corresponding Lagrangian pressure perturbation $\Delta P$ is shown in figure~\ref{delp_NS}. Equation~\ref{eq_bc_2} remains continuous across the core--crust interface, and $\Delta P$ varies smoothly throughout the star, vanishing at the stellar surface in accordance with the boundary conditions.

The stellar masses shown in these figures are chosen such that, for the same central densities and even for the weakest antikaon optical potential considered ($U_{\bar{K}} = -120$ MeV), both antikaon condensates appear in the stellar core. An exception arises for the IUFSU parametrization, in which only $K^-$ condensation occurs and the $\bar{K}^0$ condensate does not form even at the largest optical potential depth considered. In this case, we therefore select NS configurations with masses just below the maximum stable mass $M_{\mathrm{max}}$.

In figure~\ref{chi_antikaon}, we present the radial displacement eigenfunctions $\xi(r)$ for modes up to $n=10$ for NS models containing antikaons in the core. Once antikaons appear, the associated modification of the pressure--energy density relation alters the dynamical response of the star. In particular, the onset of antikaon condensation leads to a reduction of the adiabatic index. At the same time, the associated softening of the EoS results in more compact stellar configurations compared to purely nucleonic NSs.

Our calculations show that these effects manifest as a suppression of the radial displacement amplitude in the vicinity of the density at which the $K^{-}$ condensate sets in. For the relatively soft IUFSU EoS, this suppression is weaker than for the other parametrizations. The onset of $K^{-}$ condensation is therefore marked by a sharp kink in $\xi(r)$, consistent with the behavior expected from equation~\ref{eq_bc_2}. A similar, additional change in the slope of $\xi(r)$ is observed at the appearance of the $\bar{K}^0$ condensate. However, for the shallow optical potential $U_{\bar{K}}=-120$ MeV, this second feature is barely discernible.

Stronger antikaon attraction facilitates earlier condensation, causing both kinks to appear at larger radii, corresponding to lower local densities; as a direct consequence, the kink associated with the onset of the $\bar{K}^0$ condensate becomes progressively more pronounced with increasing optical potential depth. The corresponding Lagrangian pressure perturbation $\Delta P$ is shown in figure~\ref{delp_antikaon}; in all cases, its amplitude decreases smoothly with radius and vanishes at the stellar surface.

\section{Summary}
\label{sec:4}

In this work, we have systematically investigated the influence of antikaon condensates on the structural and dynamical properties of NSs within the framework of RMF theory. Both NL and DD-RMF parameterizations were employed to construct NS models composed of nucleons and leptons, and subsequently extended to include negatively charged and neutral antikaons ($K^-$ and $\bar{K}^0$) for a range of antikaon optical potential depths. This allowed us to explore the model dependence of antikaon condensation and its impact on the EoS at supranuclear densities \citep{Pal:2000pb, Banik:2001yw}.

We showed that antikaon condensation leads to a pronounced softening of the EoS, with the onset density and character of the phase transition governed by the underlying RMF parametrization and the strength of the antikaon--nucleon interaction. While most EoSs exhibit a continuous (second-order) transition to the kaon-condensed phase, the DDME$\delta$ parametrization displays a first-order transition for sufficiently deep optical potentials. The appearance of antikaons significantly modifies the particle composition, suppressing lepton populations and enhancing isospin symmetry in dense matter \citep{Pal:2000pb, MESQUITA2010}.

These changes in the EoS are reflected in the global properties of NSs. In particular, antikaon-induced softening reduces the maximum mass and radius of NSs, with increasingly attractive optical potentials leading to stronger reductions \citep{Char:2014cja, Banik:2001yw}. We identified parameter ranges in which the $\sim 2~M_\odot$ observational constraint from massive pulsars remains satisfied, thereby placing stringent bounds on the allowed strength of antikaon interactions in NS interiors.

We further investigate the impact of antikaon condensation on radial oscillations, thereby extending earlier studies in the literature \citep{Haensel:1989, Gondek:1997fd, Gondek:1999ad, Sagun:2020qvc}. The fundamental and higher-order radial mode frequencies are computed over the entire mass range of radially stable NS configurations. We find that the onset of $K^-$ and $\bar{K}^0$ condensates induces characteristic modifications in the mass--frequency relations, including an earlier vanishing of oscillation frequencies at the maximum stable mass. Furthermore, the eigenfunctions of radial displacement exhibit clear imprints of antikaon-induced softening, manifested as kinks associated with the condensate thresholds.

Overall, our results demonstrate that radial oscillations provide a sensitive probe of high-density composition and phase structure in NSs. The combined analysis of NS masses, radii, and oscillation spectra highlights the potential of future multi-messenger and high-frequency gravitational-wave observations to constrain antikaon condensation and the microphysics of dense matter.

%\section*{Acknowledgment}

%\bibliographystyle{elsarticle-num}
\newpage
\bibliographystyle{jcap}
\bibliography{refe}

\end{document}